%% file: 2019_nsk.tex
\documentclass[preprint,review,12pt]{elsarticle}

\include{packages}
\include{tikz_settings}

\include{commands}

\journal{Journal of Computational Physics}

\begin{document}

\begin{frontmatter}



\title{A Parabolic Relaxation Model for the Navier-Stokes-Korteweg Equations}


\author[label1]{Timon Hitz}
\ead{hitz@iag.uni-stuttgart.de}
\author[label1]{Jens Keim}
\author[label1]{Claus-Dieter Munz}
\author[label2]{Christian Rohde}

\address[label1]{Institute of Aerodynamics and Gasdynamics, University of Stuttgart, 70569 Stuttgart, Germany}
\address[label2]{Institute of Applied Analysis and Numerical Simulation, University of Stuttgart, 70569 Stuttgart, Germany}

\begin{abstract}
The isothermal Navier-Stokes-Korteweg system is a classical diffuse interface model for compressible two-phase flow.
However, the numerical solution faces two major challenges:
due to a third-order dispersion contribution in the momentum equations, extended numerical stencils are required for the flux calculation.
Furthermore, the equation of state given by  a Van-der-Waals law, exhibits non-monotone behaviour in the two-phase state space leading to imaginary eigenvalues of the Jacobian of the first-order fluxes.\\
In this work a  lower-order parabolic relaxation model is used to approximate solutions of the classical NSK equations.
Whereas an additional parabolic evolution equation for the relaxation variable has to be solved, the system involves no differential operator of higher as  second order. The use of a modified pressure function guarantees that the first-order fluxes remain hyperbolic.
Altogether, the relaxation system  is directly accessible for   standard compressible flow solvers.\\
We use the higher-order Discontinuous Galerkin spectral element method as realized in the  open source code \flexi.
The relaxation model is validated against solutions of the original NSK model for a variety of 1D and 2D test cases.
Three-dimensional simulations of head-on droplet collisions  for a range of different  collision Weber numbers underline the 
capability of the approach.
\end{abstract}

\begin{keyword}



Compressible flow with phase transition \sep
diffuse interface model \sep
isothermal Navier-Stokes-Korteweg equations \sep
Discontinuous Galerkin method

\end{keyword}

\end{frontmatter}



\section{Introduction}
\label{sec:introduction}

In fluid dynamics, the Navier-Stokes Equations (NSE) are the widely accepted model to describe the viscous motion of a single-phase fluid.
The extension to two-phase flows is a difficult issue that requires additional modelling of the phase interface.
Two fundamentally different approaches to describe interface dynamics exist, sharp interface and diffuse interface concepts.
In the sharp interface approach \cite{ishii_thermo-fluid_2011}, the computational domain is partitioned  by  codimensional manifolds   into distinct subdomains that contain only bulk fluids,  i.e.~either in the  vapour or the liquid phase.
The flow dynamics in  the  bulk domains  is described by the NSE while the solution, e.g.~the density, is discontinuous across the phase boundary.
Consequently, suitable transmission conditions have to be applied to couple the subdomains across the boundaries.
Due to this coupling, the position and time evolution of the phase interface is unknown a priori, rendering the overall approach to be a free boundary value problem.
Sharp interface methods are mainly applied to simulate droplet dynamics or large scale phenomena like liquid jet injection or when the resolution of the phase interface is computationally not feasible.
Diffuse interface methods \cite{anderson_diffuse-interface_1998} consider a finite thickness of the phase interface.
The free boundary of the phase interface is replaced by a continuous variation of the solution in the bulk phases with an appropriate order parameter.
To ensure physically sound profiles in the interfacial region, capillary stresses have to be modelled, accordingly.
The main application of diffuse interface methods are problems where the characteristic length scale of the problem is comparable to the interface thickness.
Examples include micro flows, topology changes of the interface (e.g. coalescence or break up), or transcritical flows in the vicinity of the critical point.

A well-known diffuse interface model is the Navier-Stokes-Korteweg (NSK) model.
It is based on the gradient theory of Van-der-Waals \cite{van_der_waals_thermodynamische_1894} for static equilibria.  There, an additional term  in the 
free energy functional that depends on the density gradient  
accounts for capillary effects.  In combination with a non-convex  bulk energy  this ansatz allows for two-phase equilibria. 
From this bulk energy the  Van-der-Waals (VdW) equation of state (EoS) is derived which gives a   non-monotonous relation between pressure and density.
The  modelling of dynamical two-phase flow  can then  be  tracked back to Korteweg \cite{korteweg_sur_1901} who introduced a class of models extending the 
NSE for one-phase flow.  Here we refer to the thermodynamically consistent form  of the NSK model as  employed by Dunn and Serrin \cite{dunn_thermomechanics_1986}, see also \cite{anderson_diffuse-interface_1998}.
The modelling of capillarity effects are inherently included in the NSK model  but the consistency with the gradient theory of Van-der-Waals
enforces a  third-order term in the momentum equations. 

The NSK model has been  studied extensively from the analytical point of view, see e.g. \cite{bresch_compressible_2003,hattori_solutions_1994,kotschote_strong_2008,rohde_local_2005}.
Likewise, there is  a large variety of  contributions on the numerical discretization, e.g. \cite{braack_stable_2013,coquel_sharp_2005,diehl_higher_2007,diehl_numerical_2016,gelissen_simulations_2018,giesselmann_energy_2014,gomez_isogeometric_2010,haink_local_2008,jamet_second_2001,martinez_high-order_2019,tian_local_2015}.
Numerical work on the NSK model has to face two fundamental issues which are not present in the NSE.
As outlined above, the momentum equations of the NSK model become third-order diffusion-dispersion equations which require some effort in discretizing.
Furthermore, due to the non-monotonous pressure function of the VdW EoS, the first-order part is of mixed hyperbolic-elliptic type.
Therefore, classical upwind based discretization schemes that rely on the solution of a hyperbolic Riemann problem cannot be used straightforwardly.

There are several ways to overcome these shortcomings.
In \cite{rohde_local_2010} a relaxation system has been proposed that treats the capillarity effects by a local and low-order differential operator.
The system is of second order with an additional relaxation parameter as unknown that has to fulfil a linear screened Poisson equation.
The system was further analyzed and solved numerically by Neusser \etal\cite{neusser_relaxation_2015} using a local Discontinuous Galerkin (LDG) method.
However, the elliptic constraint renders the overall system of mixed type hindering a consistent numerical approach.

In this work, a numerically more convenient  relaxation system is considered.  Motivated by  the work in \cite{corli_parabolic_2014,rohde_fully_2018} 
on scalar model problems,
a time dependent operator is used for the additional relaxation equation replacing the  Poisson equation by a parabolic evolution.
The relaxation model is parametrized by the Korteweg parameter such that if it tends to infinity, the original NSK model is formally recovered \cite{rohde_local_2010,neusser_relaxation_2015,rohde_fully_2018}.
For a fixed parameter, a modified pressure can be introduced that guarantees hyperbolicity of the convective fluxes if the Korteweg parameter is large enough. 
Thus, the straightforward use of upwind based numerical schemes is possible.  We introduce 
the new class of  relaxation models in  \cref{sec:fundamentals_diffuseinterface}  which also contains  a review  of the thermodynamic setting and 
   the original NSK model.

As the next step  we  present in \cref{sec:numerics_dgsem} a computational method  for  the parabolic relaxation model of \cite{rohde_local_2010} using  the open-source solver \flexi\footnote{https://www.flexi-project.org/} which is able to handle general hyperbolic/parabolic systems.
The solver is based on the Discontinuous Galerkin (DG) spectral element method (DGSEM) \cite{hindenlang_explicit_2012}, complemented by a finite volume sub-cell shock capturing method \cite{sonntag_shock_2014,sonntag_efficient_2017}.

Turning to numerical experiments  in Section \ref{sec:results} it is  first   shown  that  the solution of the original 
NSK model can be recovered by the parabolic relaxation model for one- and two-dimensional problems if the Korteweg parameter 
tends to infinity. Thus the approach can be used in this way as an approximation method for the original NSK model.
The implemented numerical method  allows on the one hand for the fine resolution of the diffuse interfacial layer by only a few mesh cells. On the other hand 
a robust capturing of small interface widths  is possible  in case of underresolved meshes which can be used for sharp interface computations 
 when  viscosity and capillarity vanish. This is not possible using numerical methods for the original NSK system.
Both issues,  on the Korteweg limit and on asymptotic robustness are addressed in Section  \ref{sec:results_nsk_validation}.\\ 
We proceed with multidimensional simulations. In Section \ref{sec:results_validation_2d} we analyze again the Korteweg limit and display the capability of the model
to deal with an ensemble of multiple droplets that might merge or vanish completely such that time-asymptotically the equilibrium 
solution is recovered.  The paper concludes with three-dimensional simulations of head-on droplet collisions in Section
\ref{sec_3D}.\\
It is an important feature of the relaxation approach that we observe  for all test cases  thermodynamically consistent discrete solutions.
This means  that  the  (discrete) total energy of the system is non-increasing over time. Up to our knowledge this has not been achieved for the classical NSK model except when using special implicit approaches \cite{GiesselmannMakridakis}. 


\section{Diffuse Interface Models}
\label{sec:fundamentals_diffuseinterface}

\subsection{Thermodynamic Setting}

\begin{figure}[h!tb]
   \centering
   \begin{subfigure}[b]{0.49\textwidth}
       \centering
      \includegraphics[width=\linewidth]{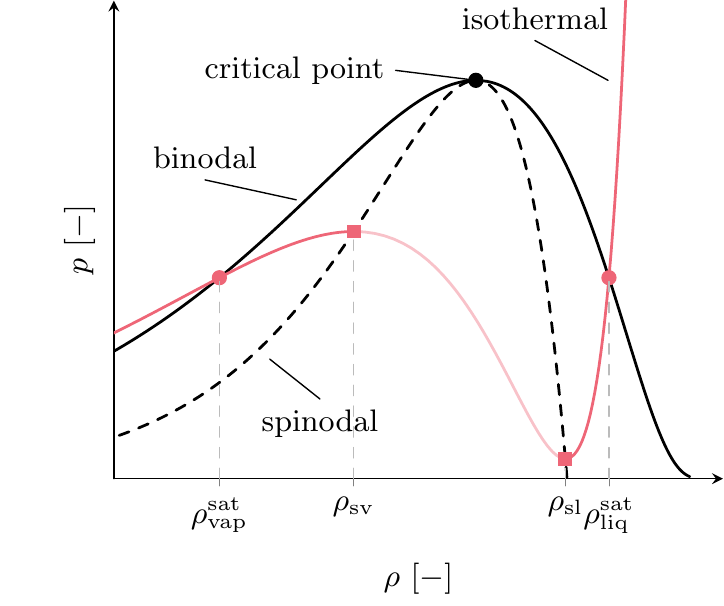}
      \label{fig:fundamentals_vdw_pressure}
   \end{subfigure}
   \begin{subfigure}[b]{0.49\textwidth}
       \centering
      \includegraphics[width=\linewidth]{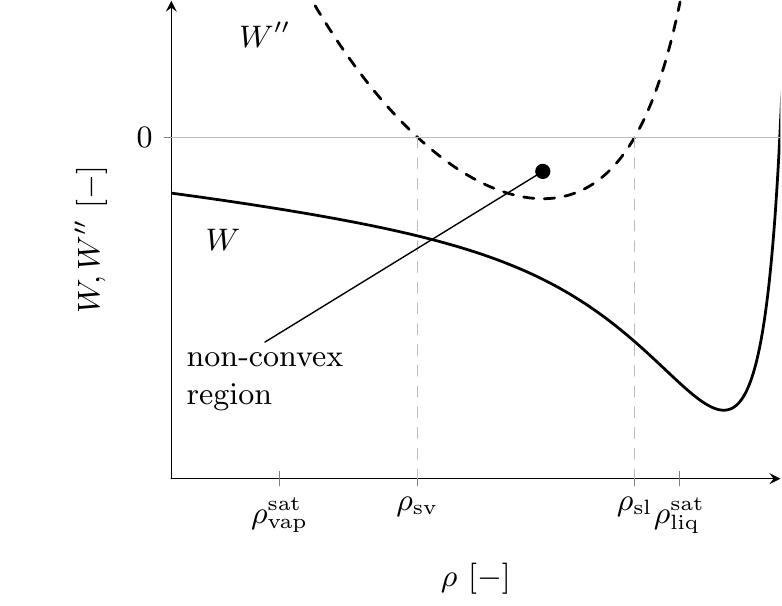}
      \label{fig:fundamentals_vdw_helmholtz}
   \end{subfigure}
   \caption{
   Graphs of the pressure $p$ (left) and Helmholtz free energy density $W(\rho)$ and its second derivative $W''(\rho)$ (right).
   }
   \label{fig:fundamentals_vdw_pressure_helmholtz}
\end{figure}

Let $\Omega \subset \R^d$ with $d \in \N$ be a bounded domain with  boundary $\partial \Omega$.
The domain is occupied by a homogeneous fluid with density $\rho(\x,t)>0$ at a uniform temperature $T \equiv T_{\mathrm{ref}}>0$.
The fluid is assumed to appear in a liquid and a vapour phase.
The two-phase structure is modelled by the Van-der-Waals (VdW) ansatz.
Then, its Helmholtz free energy density is given as
\begin{equation}
\label{eq:fundamentals_nsk_vdw_freeenergy}
W(\rho) = \rho \Rvdw T_{\mathrm{ref}} \mathrm{ln} \l( \frac{\rho}{1-\bvdw\rho} \r) - \avdw \rho^2
,
\end{equation}
where $\avdw,\bvdw,\Rvdw>0$ are material parameters.
It represents a thermodynamic potential which can be used to derive other thermodynamic quantities by, e.g., the Clausius-Duhem relation.
The pressure is consequently given by
\begin{align}
\label{eq:fundamentals_nsk_vdw_p_as_derivative}
p & = \rho W'(\rho) - W(\rho) \\
\label{eq:fundamentals_nsk_vdw_p}
& = \frac{\rho \Rvdw T_{\mathrm{ref}}}{1-\bvdw\rho} - \avdw \rho^2
.
\end{align}
For the non-dimensional VdW EoS with reduced variables (i.e. non-dimensionalized with the critical point quantities), the parameters are given by
\begin{equation}
a=3, ~b=\frac{1}{3}, ~R=\frac{8}{3}, ~T_{\mathrm{ref}}=0.85
.
\label{eq:fundamentals_vdw_parameters}
\end{equation}
This choice is kept throughout this work.
The pressure law and the Helmholtz free energy density are illustrated in \cref{fig:fundamentals_vdw_pressure_helmholtz}.
In the left panel, the coexistence curve (or binodal) is drawn.
On this curve, the fluid may exist as saturated liquid or saturated vapour, denoted by the Maxwellian densities $\rho_{\liq}^{\sat}$ and $\rho_{\vap}^{\sat}$, respectively.
These states obtain similar values for pressure and temperature, as indicated by the isothermal of the pressure law, \cref{eq:fundamentals_nsk_vdw_p} (red line).
Between the saturated states, the pressure exhibits two extrema which are denoted by $\rho_{\mathrm{sv}},\rho_{\mathrm{sl}}$.
They are called spinodal points and their locus is the spinodal curve (dashed line).
Inside the spinodal region, i.e. for density values in the interval $(\rho_{\mathrm{sv}},\rho_{\mathrm{sl}})$, the pressure function produces a so-called VdW loop that allows to connect a vapour state to a liquid state.
The loop coincides with a non-convex region of the Helmholtz free energy density.
In the right panel of \cref{fig:fundamentals_vdw_pressure_helmholtz}, the free energy density, $W(\rho)$, and its second derivative, $W''(\rho)$, are shown, where the non-convex region between the spinodal densities $\rho_{\mathrm{sv}},\rho_{\mathrm{sl}}$ exhibits a negative sign of the second derivative of $W(\rho)$.
It follows from \cref{eq:fundamentals_nsk_vdw_p_as_derivative} that the derivative of the pressure is given by $p'(\rho) = \rho W''(\rho)$, hence, the pressure is monotone decreasing in the interval $(\rho_{\mathrm{sv}},\rho_{\mathrm{sl}})$.
Due to the non-convex region of the Helmholtz free energy density, the VdW fluid is capable to describe a two-phase fluid as long as $T_{\mathrm{ref}}<T_{\mathrm{c}}$, where $T_{\mathrm{c}}$ is the temperature of the critical point.

Neglecting capillarity effects, a static equilibrium state of the liquid-vapour mixture in $\Omega$ is obtained by a minimizer of the free energy functional,
\begin{equation}
\label{eq:fundamentals_nsk_energyfunctional_minimizer_si}
\mathcal{E}_0[\rho] = \int_{\Omega} W(\rho) \d \x
.
\end{equation}
The minimizers for \cref{eq:fundamentals_nsk_energyfunctional_minimizer_si} are not unique, even for fixed total mass in $\Omega$.
However,  minimizers in the set of piecewise continuous functions take exactly two values: the Maxwellian densities of full thermodynamic equilibrium, $\rho_{\liq}^{\sat}$ and $\rho_{\vap}^{\sat}$ \cite{carr_structured_1984}.
To include capillary effects in a diffuse interface setting, the free energy functional is  extended, i.e.
\begin{equation}
\label{eq:fundamentals_nsk_energyfunctional_minimizer_di}
\mathcal{E}[\rho] = \mathcal{E}_0[\rho] + \mathcal{E}_{\Gamma}^{\mathrm{K}}[\rho]
.
\end{equation}
For the additional contribution, Van-der-Waals \cite{van_der_waals_thermodynamische_1894} proposed in his square gradient theory for liquid-vapour phase interfaces,
\begin{equation}
\label{eq:fundamentals_nsk_energyfunctional_minimizer_vdw}
\mathcal{E}_{\Gamma}^{\mathrm{K}}[\rho] = \int_{\Omega} \frac{\gammakorteweg}{2} \l|\gradient \rho \r|^2 \d \x
,
\end{equation}
where the capillarity coefficient $\gammakorteweg>0$ is a constant material parameter.
This functional penalizes the occurrence of phase interfaces by the density gradient.
Preserving total mass, the minimizers of \cref{eq:fundamentals_nsk_energyfunctional_minimizer_vdw} are smooth and uniquely determined functions.
A sequence of these diffuse interfaces selects, in the limit $\gammakorteweg \rightarrow 0$, physically relevant minimizers of \cref{eq:fundamentals_nsk_energyfunctional_minimizer_si}.

\subsection{Navier-Stokes-Korteweg Equations}

Using the thermodynamic principles of rational mechanics under the constraint of mass and momentum conservation, a stress tensor for the dynamic system can be derived, cf. \cite{anderson_diffuse-interface_1998,rohde_local_2010}, resulting in the Navier-Stokes-Korteweg (NSK) equations.
In the isothermal case for $T \equiv T_{\mathrm{ref}}$ they are
\begin{align}   
   \label{eq:fundamentals_nsk_1}
   \rho_t  + \div \l( \rho \v \r) & = 0 
   , \\
   \label{eq:fundamentals_nsk_2}
   \l( \rho \v \r)_t  + \div \l( \rho \v \otimes \v + p \I \r) & = \div \stress + \div \stresskorteweg
   ,
\end{align}
where $\v(\x,t)=(v_1(\x,t),\ldots,v_d(\x,t))\transpose$ is the velocity, $\stress\in\R^{d\times d}$ is the viscous stress tensor and $\stresskorteweg\in\R^{d\times d}$ is the Korteweg stress tensor derived from \cref{eq:fundamentals_nsk_energyfunctional_minimizer_vdw}.
The stress tensors are given by
\begin{align}
\label{eq:fundamentals_nsk_stress_nse}
\stress & = \mu \l( \gradient \v + \l(\gradient \v\r)\transpose - \frac{2}{3} \div \v \I \r)
, \\
\label{eq:fundamentals_nsk_stress_nsk}
\stresskorteweg & = \gammakorteweg \l( \rho \laplace \rho + \half \l|\gradient \rho \r|^2 \r) \I - \gammakorteweg \gradient \rho \otimes \gradient \rho
,
\end{align}
with viscosity $\mu>0$, capillarity $\gammakorteweg>0$ assumed to be constant, and the unit tensor $\I$.
The divergence of the Korteweg stress tensor can also be expressed in simplified, albeit non-conservative, form \cite{diehl_higher_2007} as
\begin{equation}
\label{eq:fundamentals_nsk_stress_nonconservative}
\div \stresskorteweg = \gammakorteweg \rho \gradient \laplace \rho
.
\end{equation}
The closure relation for $p$ is given by the isothermal VdW pressure law, \cref{eq:fundamentals_nsk_vdw_p}.
An energy functional for the dynamic system, \cref{eq:fundamentals_nsk_1,eq:fundamentals_nsk_2}, was given by Anderson \etal\cite{anderson_diffuse-interface_1998} as an extension of \cref{eq:fundamentals_nsk_energyfunctional_minimizer_di} by adding the kinetic energy contribution, i.e.
\begin{equation}
\label{eq:fundamentals_nsk_original_energy_dynamic}
\mathcal{E}[\rho] = \int_{\Omega} \l( \half \rho \l| \v \r|^2 + W(\rho) + \half \gammakorteweg \l| \gradient \rho \r|^2 \r) \d \x
.
\end{equation}

The NSK equations, \cref{eq:fundamentals_nsk_1,eq:fundamentals_nsk_2}, are non-dimensionalized in terms of reference values for density $\rho_{\mathrm{ref}}$, time $t_{\mathrm{ref}}$, length $L_{\mathrm{ref}}$, and velocity $v_{\mathrm{ref}}=\nicefrac{L_{\mathrm{ref}}}{t_{\mathrm{ref}}}$ \cite{dreyer_asymptotic_2012,diehl_higher_2007}.
The variables are given in reduced form by
\begin{equation}
\label{eq:fundamentals_nsk_nondim_vars}
\x = L_{\mathrm{ref}} \hat{\x}, \quad
t = t_{\mathrm{ref}} \hat{t}, \quad
\v = v_{\mathrm{ref}} \hat{\v}, \quad
\rho = \rho_{\mathrm{ref}} \hat{\rho}
,
\end{equation}
where the hat indicates non-dimensional variables.
The non-dimensional NSK equations read
\begin{align}   
   \label{eq:fundamentals_nsk_nondim_V1_1}
   \hat{\rho}_{\hat{t}}  + \hat{\nabla} \cdot \l( \hat{\rho} \hat{\v} \r) & = 0 
   , \\
   \label{eq:fundamentals_nsk_nondim_V1_2}
   \l( \hat{\rho} \hat{\v} \r)_{\hat{t}}  + \hat{\nabla} \cdot \l( \hat{\rho} \hat{\v} \otimes \hat{\v} + \hat{p} \I \r) & = \hat{\nabla} \cdot \hat{\stress} + \hat{\nabla} \cdot \hat{\stress}_{\mathrm{K}}
   ,
\end{align}
with the non-dimensional stress tensors expressed by
\begin{equation}
\label{eq:fundamentals_stresstensor_nsk_nondim}
  \hat{\stress} = \oneover{\mathrm{Re}} \stress
  \quad \text{and} \quad
  \hat{\stress}_{\mathrm{K}} = \oneover{\We} \stresskorteweg
  .
\end{equation}
The non-dimensional coefficients are the Reynolds number, $\mathrm{Re}$, and the Weber number, $\We$,
\begin{equation}
\label{eq:fundamentals_reynolds_weber}
\mathrm{Re} = \frac{\rho_{\mathrm{ref}} v_{\mathrm{ref}} L_{\mathrm{ref}}}{\mu} \quad \text{and} \quad \We = \frac{\rho_{\mathrm{ref}} v_{\mathrm{ref}}^2 L_{\mathrm{ref}}}{\gammakorteweg}
.
\end{equation}
Following, e.g., Dreyer \etal\cite{dreyer_asymptotic_2012}, these numbers are scaled in terms of a small parameter $\epsilonkorteweg>0$, such that
\begin{equation}
\label{eq:fundamentals_nsk_scaled_re_we}
\oneover{\mathrm{Re}} = \epsilonkorteweg, \quad
\oneover{\We} = \epsilonkorteweg^2 \hat{\gammakorteweg}
,
\end{equation}
where $\hat{\gammakorteweg}>0$ is a non-dimensional expression for the capillarity coefficient.
This scaling is similar to the choices made in \cite{rohde_local_2010,neusser_relaxation_2015} and allows to recover an appropriate asymptotic sharp interface limit for $\epsilonkorteweg \rightarrow 0$.
In the following, the hat is omitted as only non-dimensionalized equations and variables are considered.
The momentum equation, \cref{eq:fundamentals_nsk_nondim_V1_2}, is a third-order convection-diffusion-dispersion equation with mixed hyperbolic and parabolic fluxes.
The eigenvalues of the Jacobian of the first-order fluxes are given by
\begin{equation}
\label{eq:fundamentals_nsk_eigenvalues}
\lambda_1 = \v \cdot \mathbf{n} - \sqrt{p'(\rho)}
, \quad
\lambda_{2,\ldots,d+1} = \v \cdot \mathbf{n}
, \quad
\lambda_{d+2} = \v \cdot \mathbf{n} + \sqrt{p'(\rho)}
,
\end{equation}
for an arbitrary normal vector $\mathbf{n}\in\R^d$.
Due to the non-convexity of the Helmholtz free energy density, \cref{eq:fundamentals_nsk_vdw_freeenergy}, the pressure derivative becomes negative in the interval between the spinodal densities, $(\rho_{\mathrm{sv}},\rho_{\mathrm{sl}})$ for $T<T_{\mathrm{c}}$.
Hence, the eigenvalues of the convective flux Jacobian can be imaginary numbers such that the first-order part is of mixed hyperbolic-elliptic type.
The loss of hyperbolicity prevents the straightforward use of any kind of upwind schemes that use Riemann solvers, as used in FV and DG methods.
Furthermore, the third-order term of the Korteweg stress requires the calculation of additional gradients, which is computationally expensive.
Thus, some standard approaches are restricted to the use of higher-order methods complemented by simple numerical flux functions that are independent of the local wave speeds, such as the global Lax-Friedrichs (LF) flux \cite{tian_local_2015} and the central flux \cite{diehl_numerical_2016,gelissen_simulations_2018}.
Furthermore, it is mentioned that any explicit time-stepping method requires an estimate on the expected wave speeds which becomes impossible for $\lambda_{1,\ldots,d}$ being complex numbers.
\subsection{The Parabolic Relaxation Model}
\label{sec:parabolic_relaxation_mode}

To avoid these numerical challenges, Rohde \cite{rohde_local_2010} proposed a relaxation system that treats the capillarity effects by a local and low-order differential operator.
Therefore, an additional scalar field $\ckorteweg$ is introduced as a relaxation variable which satisfies a linear screened Poisson equation.
Additionally, a model parameter, $\alphakorteweg>0$, describes the asymptotic approximation towards the original equation system.
The model was investigated numerically by Neusser \etal\cite{neusser_relaxation_2015}, who showed that for $\alphakorteweg \rightarrow \infty$ the solution of the relaxation system converges to the solution of the original system.
Furthermore, they showed that the model is consistent with the first and second laws of thermodynamics and performed numerical investigations in the sharp interface limit.
Since the relaxation model of \cite{neusser_relaxation_2015,rohde_local_2010} adds an elliptical constraint of a screened Poisson equation, the numerical solution requires different discretization techniques for flow and constraint equations.

We propose an alternative form that is motivated by the analysis of some scalar model, originally discussed by Corli \etal\cite{corli_parabolic_2014}.
A time dependent operator for the additional relaxation equation is added in the form of a linear screened heat equation.
The parabolic relaxation model then reads
\begin{align}   
   \label{eq:fundamentals_nsk_relaxation_parabolic_1}
   \rho_t^{\alphakorteweg}  + \div \l( \rho^{\alphakorteweg} \v^{\alphakorteweg} \r) & = 0 ,\\
   \label{eq:fundamentals_nsk_relaxation_parabolic_2}
   \l( \rho^{\alphakorteweg} \v^{\alphakorteweg} \r)_t  + \div \l( \rho^{\alphakorteweg} \v^{\alphakorteweg} \otimes \v^{\alphakorteweg} + p^{\alphakorteweg} \I \r) & = \div \stress^{\alphakorteweg} + 
   \alphakorteweg \rho^{\alphakorteweg} \gradient \l( \ckorteweg^{\alphakorteweg} - \rho^{\alphakorteweg} \r) ,\\
   \label{eq:fundamentals_nsk_relaxation_parabolic_3}
   \betakorteweg \ckorteweg^{\alphakorteweg}_t - \epsilonkorteweg^2 \gammakorteweg \laplace \ckorteweg^{\alphakorteweg} & = \alphakorteweg \l( \rho^{\alphakorteweg} - \ckorteweg^{\alphakorteweg} \r)
   .
\end{align}
The system is of mixed parabolic-hyperbolic type.
For $\alphakorteweg \rightarrow \infty$ and $\betakorteweg\rightarrow 0$, the solution of the parabolic relaxation model approaches the solution of the original NSK equations, i.e. $(\rho^{\alphakorteweg},\v^{\alphakorteweg}) \rightarrow (\rho,\v)$.
Corli \etal\cite{corli_parabolic_2014} proposed scalings for $\betakorteweg$ by analyzing a simplified model equation.
For fixed $\epsilonkorteweg$, the parameter scales with
\begin{equation}
\label{eq:fundamentals_nsk_beta_fixede}
\betakorteweg = \betakorteweg \l(\alphakorteweg\r) = \mathcal{O} \l(\alphakorteweg^{-1}\r) 
\quad \text{for} \quad
\alphakorteweg \rightarrow \infty
,
\end{equation}
and for fixed $\alphakorteweg$, the parameter scales with
\begin{equation}
\label{eq:fundamentals_nsk_beta_fixeda}
\betakorteweg = \betakorteweg \l(\epsilonkorteweg\r) = \mathcal{O} \l(\epsilonkorteweg\r) 
\quad \text{for} \quad
\epsilonkorteweg \rightarrow 0
.
\end{equation}
The energy functional of the parabolic relaxation model has to be adjusted to account for the relaxation variable.
The correct choice of the capillary energy is similar to \cite{neusser_relaxation_2015,rohde_local_2010}, given by
\begin{equation}
\label{eq:fundamentals_nsk_energyfunctional_minimizer_pararelax_elliptical}
\mathcal{E}_{\Gamma}^{\mathrm{K}}[\rho] = 
\int_{\Omega} \l( 
\frac{\alphakorteweg}{2} \l( \rho-\ckorteweg \r)^2 + \frac{\gammakorteweg \epsilonkorteweg^2}{2}  \l| \gradient \ckorteweg \r|^2
\r) \d \x
.
\end{equation}
The total energy then becomes
\begin{equation}
\label{eq:fundamentals_nsk_pararelax_elliptical_energy_dynamic}
\mathcal{E}^{\alphakorteweg}[\rho] = \int_{\Omega} \l( 
\half \rho \l| \v \r|^2 + W(\rho) + \frac{\alphakorteweg}{2} \l( \rho-\ckorteweg \r)^2 + \frac{\gammakorteweg \epsilonkorteweg^2}{2} \l| \gradient \ckorteweg \r|^2 
\r) \d \x
.
\end{equation}

\begin{remark}[Korteweg limit]
For $d=1$, the parabolic relaxation model is
\begin{align}   
   \label{eq:fundamentals_nsk_model_1}
   \rho_t  + \l( \rho v \r)_x & = 0 ,\\
   \label{eq:fundamentals_nsk_model_2}
   \l( \rho v \r)_t  + \l( \rho v^2 + p \r)_x & = \frac{4}{3}\epsilonkorteweg v_{xx} + \alphakorteweg \rho \l( \ckorteweg - \rho \r)_x , \\
   \label{eq:fundamentals_nsk_model_3}
   \betakorteweg \ckorteweg_t - \epsilonkorteweg^2 \gammakorteweg \ckorteweg_{xx} & = \alphakorteweg \l( \rho - \ckorteweg \r)
   .
\end{align}
Since $\alphakorteweg>0$, \cref{eq:fundamentals_nsk_model_3} can be written as
\begin{equation}
   \label{eq:blub}
   \frac{\betakorteweg}{\alphakorteweg} \ckorteweg_t - \frac{\epsilonkorteweg^2 \gammakorteweg}{\alphakorteweg} \ckorteweg_{xx} = \l( \rho - \ckorteweg \r)
   .
\end{equation}
For $\alphakorteweg \rightarrow \infty$, the scaling parameter $\betakorteweg$ is given by \cref{eq:fundamentals_nsk_beta_fixeda} such that $\betakorteweg / \alphakorteweg \rightarrow 0$. 
Formally, \cref{eq:blub} then becomes
\begin{equation}
  \frac{\epsilonkorteweg^2 \gammakorteweg}{\alphakorteweg} \ckorteweg_{xx} = \l( \ckorteweg - \rho \r)
  .
\end{equation}
Inserting this equation in \cref{eq:fundamentals_nsk_model_2} yields
\begin{equation}
   \l( \rho v \r)_t  + \l( \rho v^2 + p \r)_x  = 
   \frac{4}{3}\epsilonkorteweg v_{xx} + \epsilonkorteweg^2 \gammakorteweg \rho \l(  \ckorteweg_{xx} \r)_x
   .
\end{equation}
Assuming that $(\rho - \ckorteweg) \rightarrow 0$ for $\alphakorteweg \rightarrow \infty$ and, since $\ckorteweg$ is the solution of a parabolic evolution equation, in a strong sense $(\rho_{xx} -  \ckorteweg_{xx}) \rightarrow 0$, it follows
\begin{equation}
   \l( \rho v \r)_t  + \l( \rho v^2 + p \r)_x  = 
   \frac{4}{3}\epsilonkorteweg v_{xx} + \epsilonkorteweg^2 \gammakorteweg \rho \rho_{xxx}
   ,
\end{equation}
which is the original NSK momentum \cref{eq:fundamentals_nsk_2} with the non-conservative form of the Korteweg stress.
\end{remark}

Unlike the  elliptically constrained model  the new parabolic relaxation model is in a form that is suitable for a monolithic approach by  many compressible flow solvers.
To be precise, the model can be written as
\begin{equation}
\label{eq:fundamentals_nsk_parabolic_relaxation_fluxform}
   \U_t + \div \Fbu = \sourceu
,
\end{equation}
with the vector of unknowns $\U=\l( \rho,\rho\v,\ckorteweg\r)\transpose$.
The flux vector, $\Fbu=\Fbc\l(\U\r) - \Fbv\l(\U,\gradient \U\r)$, is composed of the convective fluxes $\Fbc=(\Fb^{\mathrm{c},1},\ldots,\Fb^{\mathrm{c},d})$ and the viscous fluxes $\Fbv=(\Fb^{\mathrm{v},1},\ldots,\Fb^{\mathrm{v},d})$. Together with the source $\source$ they are given explicitly for $d=3$ by
\begin{equation}
\label{eq:fundamentals_nsk_parabolic_relaxation_fluxes_sources_model1}
\begin{split}
\Fb^{\mathrm{c},i}\l(\alphakorteweg,\betakorteweg\r) & = 
\begin{pmatrix}
\rho v_i \\ \rho v_i v_1 + \delta_{1i} p \\ \rho v_i v_2 + \delta_{2i} p \\ \rho v_i v_3 + \delta_{3i} p \\ 0
\end{pmatrix}
, \quad
\Fb^{\mathrm{v},i}\l(\alphakorteweg,\betakorteweg\r) = 
\begin{pmatrix}
0 \\ \tau_{1i} \\ \tau_{2i} \\ \tau_{3i} \\ \frac{\epsilonkorteweg^2 \gammakorteweg}{\betakorteweg} \ckorteweg_{x_i}
\end{pmatrix}
, \\
\source\l(\alphakorteweg,\betakorteweg\r)&  = 
\begin{pmatrix}
0 \\ \alphakorteweg \rho \l(\ckorteweg-\rho\r)_{x_1} \\ \alphakorteweg \rho \l(\ckorteweg-\rho\r)_{x_2} \\ \alphakorteweg \rho \l(\ckorteweg-\rho\r)_{x_3} \\ \frac{\alphakorteweg}{\betakorteweg} \l( \rho - \ckorteweg \r)
\end{pmatrix}
,
\end{split}
\end{equation}
where $\delta_{ij}$ is the Kronecker symbol.
The eigenvalues of the Jacobian of the hyperbolic flux $\Fbc$ are those from \cref{eq:fundamentals_nsk_eigenvalues}.
In the following the balance law, \cref{eq:fundamentals_nsk_parabolic_relaxation_fluxform}, together with the fluxes \cref{eq:fundamentals_nsk_parabolic_relaxation_fluxes_sources_model1}, is referred to as \emph{relaxation model I}.

However, if we define
\begin{equation}
\label{eq:fundamentals_nsk_modified_pressure}
p_{\alphakorteweg} \l(\rho,\alphakorteweg \r) = p\l(\rho \r) + \alphakorteweg \frac{\rho^2}{2}
,
\end{equation}
one can rewrite \cref{eq:fundamentals_nsk_parabolic_relaxation_fluxform,eq:fundamentals_nsk_parabolic_relaxation_fluxes_sources_model1} in an equivalent form with the choices
\begin{equation}
\label{eq:fundamentals_nsk_parabolic_relaxation_fluxes_sources_model2}
\begin{split}
\Fb^{\mathrm{c},i}\l(\alphakorteweg,\betakorteweg\r) & = 
\begin{pmatrix}
\rho v_i \\ \rho v_i v_1 + \delta_{1i} p_{\alphakorteweg} \\ \rho v_i v_2 + \delta_{2i} p_{\alphakorteweg} \\ \rho v_i v_3 + \delta_{3i} p_{\alphakorteweg} \\ 0
\end{pmatrix}
, \quad
\Fb^{\mathrm{v},i}\l(\alphakorteweg,\betakorteweg\r) = 
\begin{pmatrix}
0 \\ \tau_{1i} \\ \tau_{2i} \\ \tau_{3i} \\ \frac{\epsilonkorteweg^2 \gammakorteweg}{\betakorteweg} \ckorteweg_{x_i}
\end{pmatrix}
, \\
\source\l(\alphakorteweg,\betakorteweg\r) & = 
\begin{pmatrix}
0 \\ \alphakorteweg \rho \ckorteweg_{x_1} \\ \alphakorteweg \rho \ckorteweg_{x_2} \\ \alphakorteweg \rho \ckorteweg_{x_3} \\ \frac{\alphakorteweg}{\betakorteweg} \l( \rho - \ckorteweg \r)
\end{pmatrix}
.
\end{split}
\end{equation}
We name this model accordingly, \emph{relaxation model II}.
The eigenvalues of the Jacobian of the first-order flux are now
\begin{equation}
\label{eq:fundamentals_nsk_parabolic_eigenvalues_model2}
\lambda_1 = \v \cdot \mathbf{n} - \sqrt{p_{\alphakorteweg}'(\rho)}
, \quad
\lambda_{2,\ldots,d+1} = \v \cdot \mathbf{n}
, \quad
\lambda_{d+2} = \v \cdot \mathbf{n} + \sqrt{p_{\alphakorteweg}'(\rho)}
,
\end{equation}
for an arbitrary normal vector $\mathbf{n}\in\R^d$.
Even for densities where the Helmholtz free energy density, \cref{eq:fundamentals_nsk_vdw_freeenergy}, becomes non-convex, the flux $\Fbc\l(\alphakorteweg\r)$ remains strictly hyperbolic for large enough values for the Korteweg parameter $\alphakorteweg$ \cite{neusser_relaxation_2015},
\begin{equation}
\label{eq:fundamentals_alpha_fullyhyperbolic_limit}
\alphakorteweg > \alpha_{*} = \frac{\l| \min \l( W''(s) : s \in \l(\rho_{\mathrm{sv}},\rho_{\mathrm{sl}}\r) \r) \r| }{2}
.
\end{equation}

It was found that the \emph{relaxation model II} performs very well in the Korteweg limit for $\alphakorteweg >\!> 0$ and it therefore is the standard model used in this work.

\section{Numerical Methods}
\label{sec:numerics_dgsem}

The open source code \flexi was designed on the basis of the Discontinuous Galerkin (DG) method to enable efficient and reliable computations of compressible flow problems.
For a survey of the basic features we refer to \cite{hindenlang_explicit_2012} and for sub-cell resolution as favoured shock-capturing technique to \cite{sonntag_shock_2014,sonntag_efficient_2017}.
In this contribution, we extend the \flexi code to cover the compressible NSK systems \cref{eq:fundamentals_nsk_1,eq:fundamentals_nsk_2} and the relaxation model II, \cref{eq:fundamentals_nsk_parabolic_relaxation_fluxform,eq:fundamentals_nsk_parabolic_relaxation_fluxes_sources_model2}.
The numerical methods are outlined in this section.

\subsection{The Discontinuous Galerkin Spectral Element Method}

The computational domain $\Omega$ is divided into non-overlapping hexahedral grid cells, which are allowed to be organized fully unstructured in a conforming way.
Each grid cell is mapped onto a reference cube $E= [-1,1]^3$.
The mapping $\mathbf{x}(\xi)$ between reference space, $\xi=(\xi^1,\xi^2,\xi^3)\transpose$, and physical space, $\mathbf{x}=(x^1,x^2,x^3)\transpose$, transforms the balance equations, \cref{eq:fundamentals_nsk_parabolic_relaxation_fluxform}, into
\begin{equation}
   \jac \U_t + \divXI \Fu = \jac \source \quad \text{in } E
   ,
   \label{eq:numerics_general_conservation_law_refelem}
\end{equation}
where the transformed fluxes $\F$ are defined by
\begin{equation}
   \div \Fb = \oneover{\jac} \sum_{i=1}^3 \fracp{\jac a^i \cdot \Fb}{\xi^i} = \oneover{\jac} \divXI \F
   .
   \label{eq:numerics_transformedfluxes}
\end{equation}
The transformation is given in terms of covariant basis vectors $a^i$ and the Jacobian of the transformation $\jac$ as described by Kopriva \cite{kopriva_metric_2006,kopriva_implementing_2009} such that the mapping is free stream preserving.
In \cref{eq:numerics_general_conservation_law_refelem,eq:numerics_transformedfluxes}, the divergence operator $\divXI$ indicates the derivative with respect to directions in reference space.

The solution and each component of the contravariant fluxes $\F^m$ in the reference element are approximated by a polynomial tensor product basis of degree $N$,
\begin{align}
   \label{eq:numerics_solutionpolynomial}
   \U_h \l(\xi,t \r) & := \sum_{i,j,k=0}^{N} \hatU_{ijk}(t) \Psi_{ijk}(\xi), \\
   \label{eq:numerics_fluxpolynomial}
   \F^m_h (\U,\gradient \U ) & := \sum_{i,j,k=0}^{N} \hat{\F}^m_{ijk} \Psi_{ijk}(\xi)
   ,\quad m=1,2,3,
\end{align}
where the interpolation nodes, $\{\xi_i\}_{i=0}^N$, are the Gauss-Legendre quadrature points on the interval $[-1,1]$.
The polynomial basis is given as tensor products of one-dimensional Lagrange polynomials $\ell$ of degree $N$,
\begin{equation}
   \Psi_{ijk} = \ell_i(\xi^1) \ell_j(\xi^2) \ell_k(\xi^3)
   .
   \label{eq:numerics_polynomialsbasis}
\end{equation}
The symbol $\hatU_{ijk}(t)$ denotes the nodal degrees of freedom and $\hat{\F}^m_{ijk}$ are the corresponding nodal degrees of freedom in the $m$-th component of the flux $\F$.

The transformed balance law is multiplied by a test function, $\phi$, and integrated on the reference element $E$ to yield the variational formulation,
\begin{equation}
   \intE{\jac \U_t \phi} + \intE{\divXI \F \phi}  = \intE{\jac \source \phi}
   .
   \label{eq:numerics_variationalformulation}
\end{equation}
The weak form is obtained by integration by parts of the second integral,
\begin{equation}
   \intE{\jac \U_t \phi} - \intE{\F \cdot \gradientXI \phi} + \intpE{ \l(\F \cdot \nref \r)^* \phi }  = \intE{\jac \source \phi}
   ,
   \label{eq:numerics_weakformulation}
\end{equation}
where $\nref$ denotes the unit normal vector on the reference element face.
To obtain a DG formulation, the Galerkin method is applied to \cref{eq:numerics_weakformulation} and the test functions are chosen identical to the polynomial basis functions of the solution approximation.
Eq. \eqref{eq:numerics_solutionpolynomial} and \eqref{eq:numerics_fluxpolynomial} are inserted into the weak form, \cref{eq:numerics_weakformulation}, and the integrals are evaluated numerically using Gauss-Legendre quadrature rules.
Collocation is used by choosing the same points for integration as for interpolation.
The Kronecker delta property of the Lagrange polynomials, $\ell_i(\xi_i)=\delta_{ij}$, is utilized to yield the semi-discrete form.
Since the solution is allowed to be discontinuous across element faces, it may be double valued and the surface fluxes has to be obtained using numerical flux functions (denoted by the asterisk).
In this work, the local and global Lax-Friedrichs (LF) method as well as the approximate Harten-Lax-van Leer-contact (HLLC) Riemann solver are employed.
For a detailed description, the interested reader is referred to Toro \cite{toro_riemann_2009}.

The method described above is sufficient to discretize first-order balance laws.
In second- or third-order equations, fluxes and source terms depend on the gradients of the solution.
A common approach for DG methods is the use of lifting procedures, such as the BR1 scheme of Bassi and Rebay \cite{bassi_high-order_1997}.
For example, a second-order system of equations is rewritten into a first-order system by approximating the gradient as
\begin{equation}
\Q_1 = \gradient \U \quad \Leftrightarrow \quad \Q_1 - \gradient \U = 0.
\end{equation}
Transformation into the reference element yields the conservation equation
\begin{equation}
\Q_1 - \oneover{\jac} \divXI \mathcal{U} = 0, \quad \mathcal{U} = \l( \jac a^1,\jac a^2,\jac a^3\r)\transpose \U.
\end{equation}
It is discretized by the DGSEM method as described by Hindenlang \etal\cite{hindenlang_explicit_2012,hindenlang_mesh_2014}.
As numerical flux function, the central flux is used.
All necessary gradients are calculated directly from their primitive state variables, e.g. the velocity is lifted directly for the calculation of the viscous stresses.
This method can be extended to third-order equations, e.g. the second gradient of density of the original NSK equations is calculated by a second application of the BR1 lifting procedure,
\begin{equation}
\Q_2 = \gradient \l(\gradient \rho\r)\cdot\I \quad \Leftrightarrow \quad \Q_2 - \gradient \l(\gradient \rho \r)\cdot\I = 0.
\end{equation}
Transformation into the reference element yields the conservation equation
\begin{equation}
\Q_2 - \l(\oneover{\jac} \divXI \mathcal{U}_2\r)\cdot\I = 0, \quad \mathcal{U}_2 = \l( \jac a^1,\jac a^2,\jac a^3\r)\transpose \gradient \rho,
\end{equation}
which can also be discretized by the DGSEM method.

Using the method of lines, the solution can be advanced in time using either explicit third or fourth-order low storage Runge-Kutta (RK) methods as described by Kennedy \etal\cite{kennedy_low-storage_2000} or implicitly by a fourth-order explicit singly diagonally implicit Runge-Kutta (ESDIRK) scheme with six stages \cite{kennedy_additive_2003} as proposed by Vangelatos and Munz \cite{vangelatos_study_2018}.
The advantage of using implicit schemes is their larger time step size due to their unconditional stability.
Explicit schemes on the other hand have to fulfil the parabolic time step restriction \cite{gassner_discontinuous_2009},
\begin{equation}
   \label{eq:numerics_dfl}
   \Delta t \le \cfl~ \beta_{\mathrm{RK}}(N) \frac{\Delta \xi^2}{\lambda^{\mathrm{v}} (2 N+1)^2},
\end{equation}
where $\beta_{\mathrm{RK}}(N)$ is the viscous scaling factor for RK time integration, and $\lambda^{\mathrm{v}}$ is the maximum eigenvalue of the diffusion matrix,
\begin{equation}
\label{eq:numerics_nsk_visc_eigenvalues}
\lambda^{\mathrm{v}} = \max \l( \epsilonkorteweg,  \frac{\epsilonkorteweg^2 \gammakorteweg}{\betakorteweg} \r)
.
\end{equation}
For particularly small choices of $\betakorteweg$, the explicit time step size is dominated by the screened heat equation for the relaxation variable.
In numerical experiments, we found that for $\cfl=100$, implicit time integration was faster than explicit time integration while the solution remained similar.

Note that the source term of the parabolic relaxation model for the NSK equations, \cref{eq:fundamentals_nsk_relaxation_parabolic_1,eq:fundamentals_nsk_relaxation_parabolic_2,eq:fundamentals_nsk_relaxation_parabolic_3}, includes non-conservative products which depend on the gradient of the solution.
In this work, the straightforward implementation of a point-wise source term using the lifted gradients proved to be stable and a path-conservative discretization, as proposed in \cite{neusser_relaxation_2015}, is not required.

\subsection{Finite Volume Sub-Cell Approach}
\label{sec:numerics_fv}

As a higher-order approach, DGSEM exhibits instabilities in resolving steep gradients or even discontinuities in the solution.
Due to the Gibbs phenomenon, the approximate solution oscillates and in extreme cases, the simulation breaks off.
A shock capturing method was developed by Sonntag and Munz \cite{sonntag_shock_2014,sonntag_efficient_2017} which switches locally to a total variation diminishing (TVD) second-order Finite Volume method on sub-cells in troubled elements.
Discontinuities or steep-gradient waves of interest in this work are shock waves and in particular phase boundaries in the regime of large Reynolds and large Weber numbers.

\begin{figure}[h!tb]
   \centering
   \includegraphics[width=0.7\linewidth]{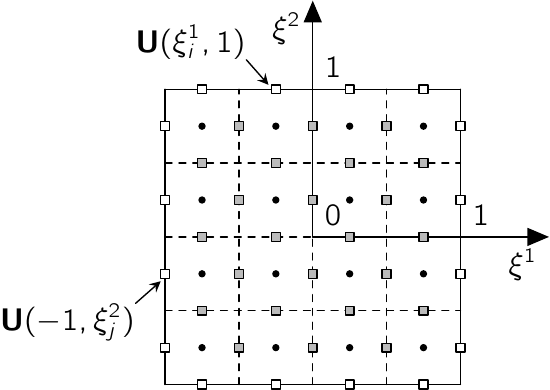}
   \caption{
   Discretization in a 2D grid cell using equidistantly distributed FV sub-cells.
   }
   \label{fig:numerics_subcells}
\end{figure}

If an oscillation is detected in a DG element, $(N+1)^3$ equidistantly distributed sub-cells are introduced as shown in \cref{fig:numerics_subcells}.
The conversion from DG polynomials to integral mean values in each sub-cell is done by numerical integration using Gauss quadratures using $N+1$ nodes for each sub-cell, hence integration is exact.

Each sub-cell is now a control volume where the FV method can be expressed by the weak form, \cref{eq:numerics_weakformulation}, with test function $\phi=1$.
The volume integral vanishes while for the surface integral the midpoint rule is employed.
As numerical flux functions of the surface integral, LF and HLLC methods are used.

In order to reduce numerical dissipation, a second-order TVD reconstruction is applied with the simple MinMod limiter.
For diffusion fluxes, Green's identity is applied together with a continuous reconstruction of the gradient, cf. \cite{blazek_computational_2005}.
Note that this particular approximation of gradients is restricted to second-order systems and, hence, the FV shock-capturing as described here cannot be applied to the original third-order NSK system.

The FV sub-cell method is only applied in elements that generate oscillations, caused by steep gradients or discontinuities.
To identify troubled elements, the indicator of Persson and Peraire \cite{persson_sub-cell_2006} is used which compares the polynomial coefficients of a representation of the solution with a hierarchical basis function.
If the solution is smooth, the coefficients (or modes) decay quickly.
If steep gradients are present, polynomials tend to oscillate and the coefficients of the higher modes do not decrease.
By user-specified threshold values, the solution in the troubled element is switched to the FV sub-cell representation.

\section{Results}
\label{sec:results}

In this section, numerical experiments of the parabolic relaxation model for the isothermal NSK equations are presented.
First, the model is validated against reference solutions from the original NSK model using 1D benchmark problems.
A one-dimensional multiphase shock tube Riemann problem is simulated to drive the parabolic relaxation model towards the sharp interface limit.
In 2D, several simulations of up to 101 droplets that merge or evaporate are presented.
Then, 3D simulations of droplet collisions are shown.

\subsection{1D Test Cases}
\label{sec:results_nsk_validation}

Numerical tests were performed to investigate the parabolic relaxation model for the NSK equations in the Korteweg limit.
First, static solutions of quiescent phase interfaces with zero velocity were used to determine the behaviour of the relaxation parameters $\alphakorteweg$, $\betakorteweg$.
A reference solution was produced by the original NSK model.
Secondly, the model capabilities in the sharp interface limit were studied.

\subsubsection{Validation of the Relaxation Model in  the Korteweg Limit}

\paragraph{Static solutions}
For $v\equiv 0$, static solutions, $\rho=\rho(x)$, for the original NSK equations can be derived.
They are used for the validation of the parabolic relaxation model.
An exact static solution is given by
\begin{align}
\label{eq:results_nsk_static_initialcondition}
\rho(x) & = \frac{\rho_{\liq}^{\sat}+\rho_{\vap}^{\sat}}{2} + \frac{\rho_{\liq}^{\sat}-\rho_{\vap}^{\sat}}{2} \mathrm{tanh} \l( \frac{x+0.4}{2\sqrt{\gammakorteweg \epsilonkorteweg^2}} \r), \\
\v(x) & = \l(0,0,0\r)\transpose
,
\end{align}
which was also used as initial condition.
\begin{figure}[!ht]
   \centering
   \includegraphics[width=\linewidth]{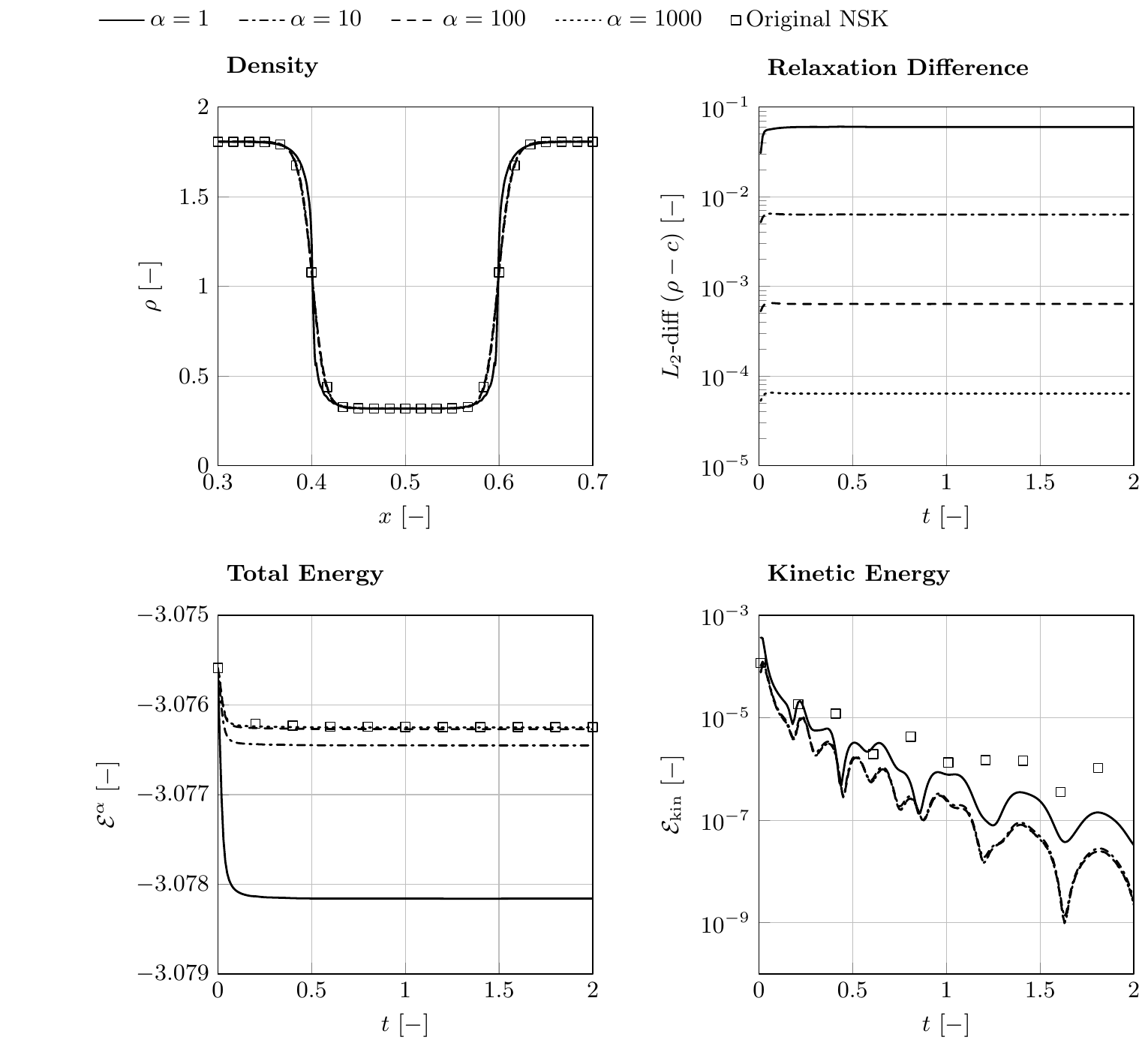}
   \caption{Results for a static phase boundary in 1D for varying $\alphakorteweg$ at $\betakorteweg=0.01$.
            }
   \label{fig:results_nsk_1d_static_alpha}
\end{figure}
For $T=0.85$, the Maxwellian densities are $\rho_{\vap}^{\sat}=0.3197$ and $\rho_{\liq}^{\sat}=1.8071$.
The capillarity parameter was $\gammakorteweg=1$ and the viscosity $\epsilonkorteweg=0.01$.
The Korteweg parameter varied between $\alphakorteweg=1,10,100,1000$ while the parabolic relaxation parameter remained constant at $\betakorteweg=0.01$.
The domain, $\Omega = (0,1)$, was discretized into 200 elements and periodic boundary conditions were employed.
The polynomial degree was $N=3$, the numerical flux was computed by the HLLC method and time integration was done implicitly with $\cfl=100$, using a
 fourth-order ESDIRK scheme with six stages.

\Cref{fig:results_nsk_1d_static_alpha} shows the solution for density at time instance $t=2.0$, the $L_2$ difference  $\rho-c$, as well as the decay of total and kinetic energies.
The results show that for $\alphakorteweg>10$ the density profiles converge to the solution of the original NSK model.
For increasing values of $\alphakorteweg$, the $L_2$ difference between the density and the relaxation variable $c$ decrease, but for all cases, it converges to a constant value.
The total energy for all cases shows a monotone decrease up to a constant value.
The kinetic energy approaches values of $\mathcal{O}(10^{-7})$ although it oscillates around a decreasing mean value.
For $\alphakorteweg>100$ the total and kinetic energies coincide.
In addition, the total energy of the relaxation model lies very close to the total energy of the original NSK model, hence, $\alphakorteweg=100$ is a sufficient choice for the Korteweg limit.

\begin{figure}[!ht]
   \centering
   \includegraphics[width=\linewidth]{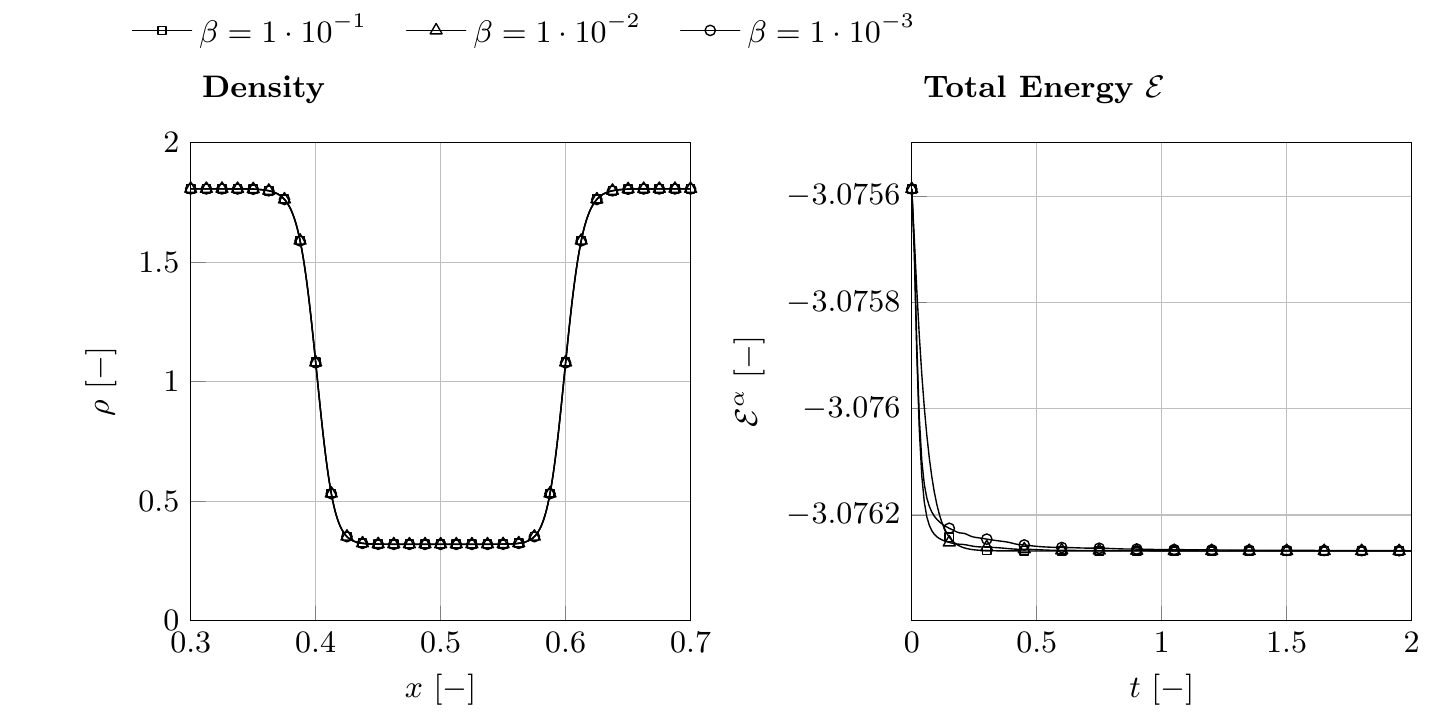}
   \caption{Results for a static phase boundary in 1D for varying $\betakorteweg$ at $\alphakorteweg=100$.
            }
   \label{fig:results_nsk_1d_static_beta}
\end{figure}

\Cref{fig:results_nsk_1d_static_beta} shows the density at $t=2.0$, as well as the total energy for $\alphakorteweg=100$ and varying $\betakorteweg=0.1,0.01,0.001$.
Independently of $\betakorteweg$, the density converges on the same solution and a monotone energy decay is established.
As discussed in \cref{sec:parabolic_relaxation_mode}, we expect to approach the original NSK model for $\alphakorteweg\rightarrow \infty$ and $\betakorteweg\rightarrow 0$.
In this case a small value of $\ckorteweg_t$ is obtained such that convergence occurs also for moderate values of $\betakorteweg$.

\begin{figure}[!ht]
   \centering
   \includegraphics[width=\linewidth]{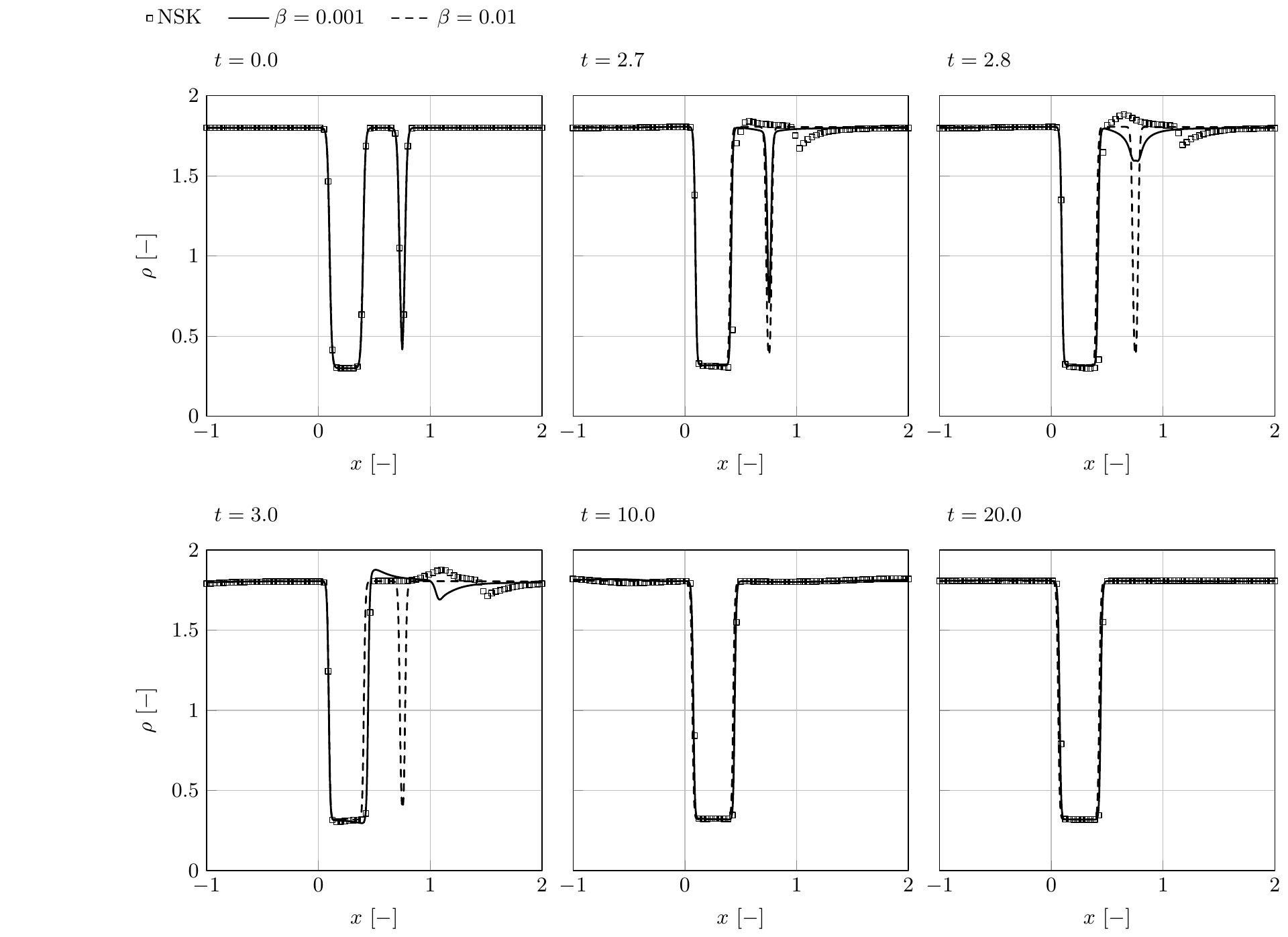}
   \caption{Solutions of density of two merging bubbles in 1D at time instances $t=0,2.7,2.8,3,10,20$.
            }
   \label{fig:results_nsk_1d_mergingbubbles}
\end{figure}

\paragraph{Merging Bubbles (Ostwald Ripening)}
In another one-dimensional test case, two bubbles were initialized in a saturated liquid phase.
The initial conditions were chosen such that the smaller bubble condensates and vanishes while transferring its mass to the larger one without coalescing directly.
They were
\begin{align}
\label{eq:results_nsk_mergingbubbles_initialcondition}
\rho(\x,t=0) & = \frac{\l( n \rho_{\vap} - \l(n-2\r) \rho_{\liq} \r)}{2} + \sum_{i=1}^{n} \frac{\rho_{\liq}-\rho_{\vap}}{2} \mathrm{tanh} \l( \frac{d_i-r_i}{2\sqrt{\gammakorteweg \epsilonkorteweg^2}} \r), \\
\v(\x,t=0) & = \l(0,0,0\r)\transpose
,
\end{align}
where $n=2$ and $d_1=x-0.25$, $d_2=x-0.75$ and $r_1=0.15$, $r_2=0.05$.
The initial densities were $\rho_{\liq}=1.8$ and $\rho_{\vap}=0.3$, the viscosity and capillarity were $\epsilonkorteweg=0.01$ and $\gammakorteweg=1$, and the Korteweg parameter was $\alphakorteweg=100$.
The second relaxation parameter, $\betakorteweg$, was varied between $0.01$ and $0.001$ to compare its influence on the time evolution towards the equilibrium state.
Therefore, the solution of the original NSK model serves as reference.
The domain, $\Omega= (-1,2)$, was discretized into $400$ elements with a polynomial degree of $N=3$.
For the relaxation model, time integration was implicit with $\cfl=100$ using a fourth-order ESDIRK scheme with six stages, and the numerical fluxes were computed by the HLLC method.
For the original NSK model, time integration was explicit using a fourth-order RK scheme with five stages and the LF method as numerical flux function.
The solution of density at various time instances is shown in \cref{fig:results_nsk_1d_mergingbubbles}.
All simulations converge to the same equilibrium state with two phase boundaries only, but time evolution at earlier times differs.
Using a smaller value for $\betakorteweg$, the equilibrium state is reached faster, but the waves emitted from the vanishing smaller bubble appear later than predicted by the original NSK model.

\paragraph{Travelling wave solution}
A travelling wave solution was simulated where a single phase boundary was propagated.
The test case was adapted from Diehl \cite{diehl_higher_2007} who suggested an exact solution of the original NSK equations using the Maxwellian densities at $T=0.85$.
The viscosity, the capillarity coefficient, and the propagation velocity were $\epsilonkorteweg=0.01366$, $\gammakorteweg=5.35918$, and $v_0=-0.3214$, respectively.
The wave was initialized by
\begin{align}
\label{eq:results_nsk_travellingwave_initialcondition}
\rho(x,t=0) & = \frac{\rho_{\liq}^{\sat}+\rho_{\vap}^{\sat}}{2} + \frac{\rho_{\liq}^{\sat}-\rho_{\vap}^{\sat}}{2} \mathrm{tanh} \l( \frac{x-0.5}{2\sqrt{\gammakorteweg \epsilonkorteweg^2}} \r), \\
\v(x,t=0) & = \l(v_0,0,0\r)\transpose
.
\end{align}
The relaxation parameters were chosen to be $\alphakorteweg=100$ and $\betakorteweg=0.01,0.001$.
The domain, $\Omega = (-1,1)$, was discretized into $400$ elements with a polynomial degree of $N=3$.
Time integration was implicitly with $\cfl=100$ using a fourth-order ESDIRK scheme with six stages, and the numerical fluxes were computed by the HLLC method.
The reference solution of the original NSK model was produced using the same spatial discretization and explicit time integration with a fourth-order RK scheme with five stages.
The numerical fluxes were computed by the LF method.

\begin{figure}[!ht]
   \centering
   \includegraphics[width=\linewidth]{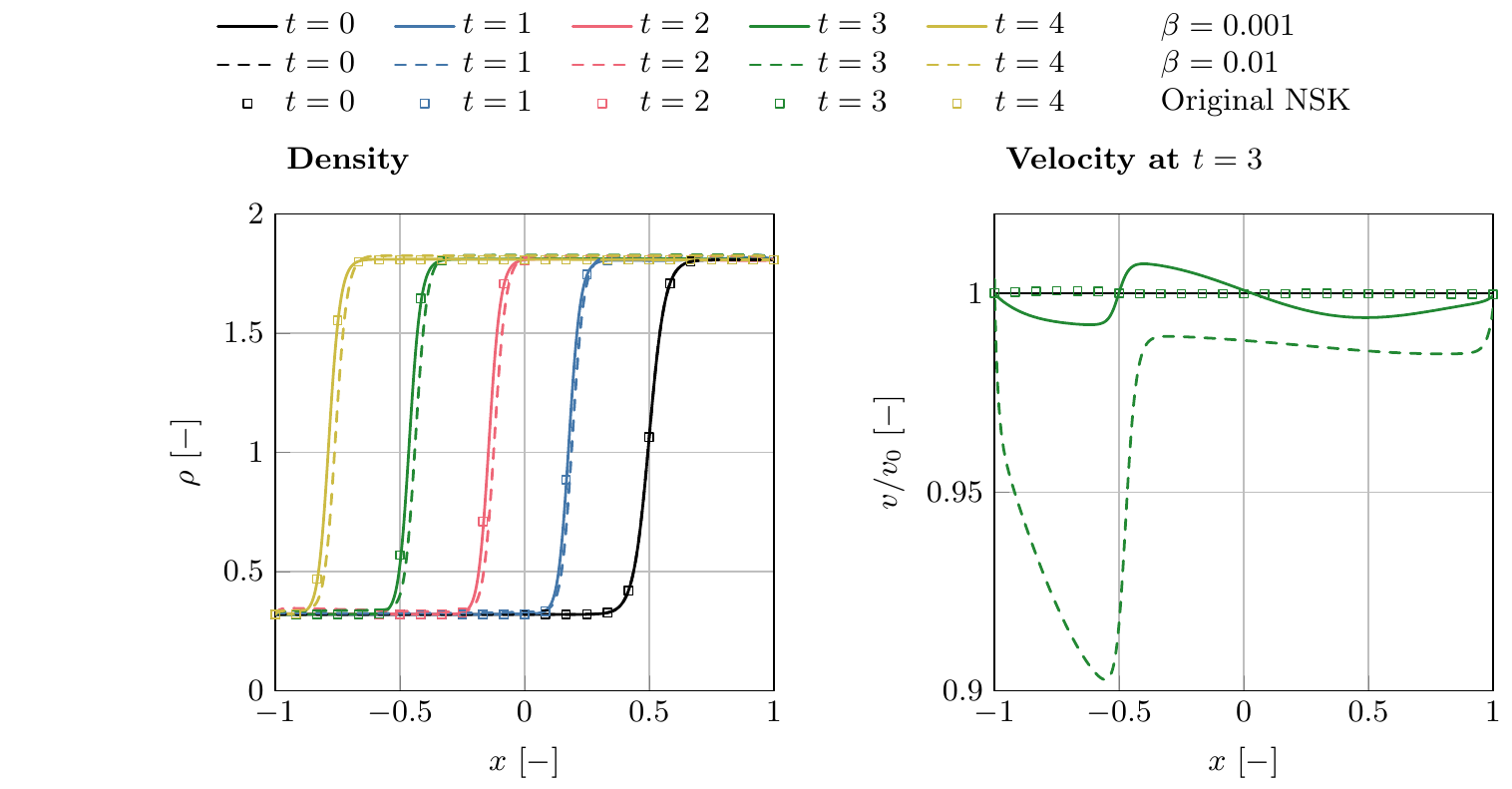}
   \caption{Solutions of density and relative velocity for a travelling wave in 1D at time instances $t=0,1,2,3,4$.
            }
   \label{fig:results_nsk_1d_travelingwave_singleinterface}
\end{figure}

\Cref{fig:results_nsk_1d_travelingwave_singleinterface} shows the results of density at time instances $t=0,1,2,3,4$ and relative velocity at $t=3$ for the relaxation model compared to results obtained from the original NSK model.
Since the initial condition is not equivalent to the travelling wave solution, for the relaxation model, pressure waves emit to adjust the density profile in the interfacial region.
This results in velocity fluctuations which decrease at later time instances.
At $t=3$, the relaxation model with $\betakorteweg=0.001$ produces a maximum of approximately $1 \si{\percent}$ velocity fluctuations while the original model hardly fluctuates at all.
For $\betakorteweg=0.01$, the density profile propagates with a different wave speed and, furthermore, the maximum velocity fluctuation reaches up to $10~\%$ at $t=3$.
For unsteady problems, it is preferable to choose a smaller value for $\betakorteweg$ than $\betakorteweg=1/\alphakorteweg$, as proposed by \cite{corli_parabolic_2014}.

\subsubsection{1D Riemann Problem in the Sharp Interface Limit}

The use of the parabolic relaxation model for the NSK diffuse interface method enables numerical simulations in the sharp interface limit.
Since the relaxation model is a second-order system of PDEs, a simple FV shock capturing method can be used to deal with strong gradients and discontinuities such as phase boundaries and shock waves.
The sharp interface is achieved as the asymptotic limit for $\epsilonkorteweg \rightarrow 0$.

Results were produced in 1D by solving a multiphase shock tube Riemann problem.
In this scenario, a stable liquid phase lies opposite of a stable vapour phase at the initial discontinuity at $x_0$.
Due to the differences in density and pressure, the solution structure of the isothermal multiphase Riemann problem develops, cf. \cite{rohde_fully_2018}, where the density profile in the interfacial region is resolved.

\paragraph{Simulation Setup}
The domain was $\Omega=(0,1)$ and the initial data were
\begin{equation}
(\rho,v)\transpose = 
\begin{cases}
(2  , 0.0 )\transpose & \quad \text{if} ~ x < 0.5, \\
(0.1, 0.0 )\transpose & \quad \text{if} ~ x > 0.5
.
\end{cases}
\label{eq:results_nsk_sharp_shocktube_initialdata}
\end{equation}
The relaxation parameters were $\alphakorteweg=100$, $\betakorteweg=0.001$, the capillary coefficient was $\gammakorteweg=1$, and the viscosity parameter varied, $\epsilonkorteweg=0.01,0.001,0.0001,0.00001$.
The domain was discretized into 200 elements and a polynomial degree of $N=3$.
FV shock capturing was used to resolve steep gradients.
Oscillations were detected by a Persson indicator with the thresholds $\mathrm{Ind}_{\mathrm{upper}}=-6$ and $\mathrm{Ind}_{\mathrm{lower}}=-8$.
The numerical flux function was the HLLC method.
The time discretization method was explicit using a fourth-order RK method with five stages and $\cfl=0.1$.
At the domain boundaries, Dirichlet boundary conditions were used.

\paragraph{Simulation Results}

\begin{figure}[!ht]
   \centering
   \includegraphics[width=\linewidth]{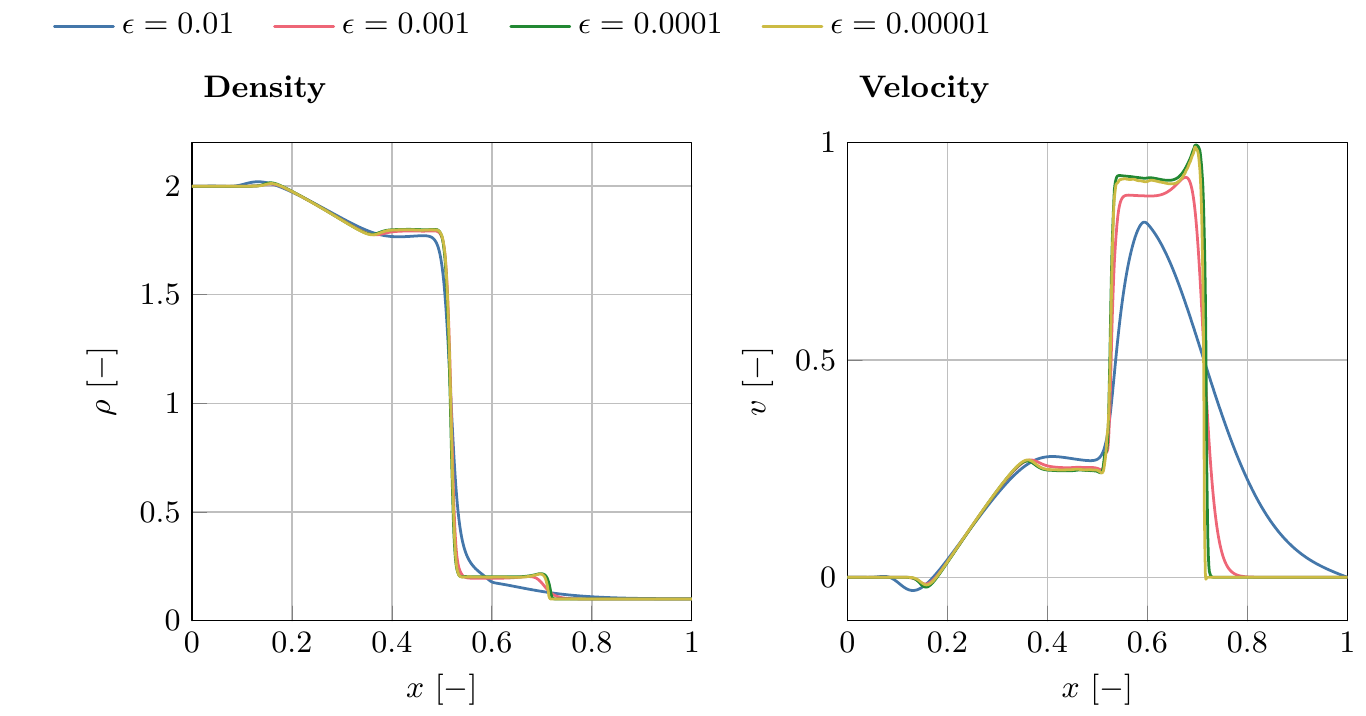}
   \caption{Density and velocity profiles of the multiphase shock tube problem at $t=0.11$ using \emph{relaxation model II}.
            }
   \label{fig:results_nsk_riemann}
\end{figure}

The density and velocity profiles of the solutions are visualized in \cref{fig:results_nsk_riemann} at $t=0.11$.
A rarefaction wave propagates to the left and expands the liquid phase.
The liquid undergoes phase transition across a phase boundary while on the far right, a shock wave propagates into the initial vapour phase.
For the largest value of $\epsilonkorteweg$, the shock wave is completely smeared out and smoothly connects to the phase boundary.
For decreasing values of $\epsilonkorteweg$, the phase boundary and shock wave sharpen until the solution converges for $\epsilonkorteweg < 0.0001$.
The thickness of the phase interface is determined by the grid resolution in the interfacial region.
Around shock, rarefaction and phase boundary waves over and undershoots are visible for $\epsilonkorteweg \leq 0.001$, especially in the velocity.
Nevertheless, the simulation remained stable in time, even for extremely low values for $\epsilonkorteweg$. This is not possible when using 
known numerical methods for the  discretization  of the classical NSK model directly.

\begin{figure}[!ht]
   \centering
   \includegraphics[width=\linewidth]{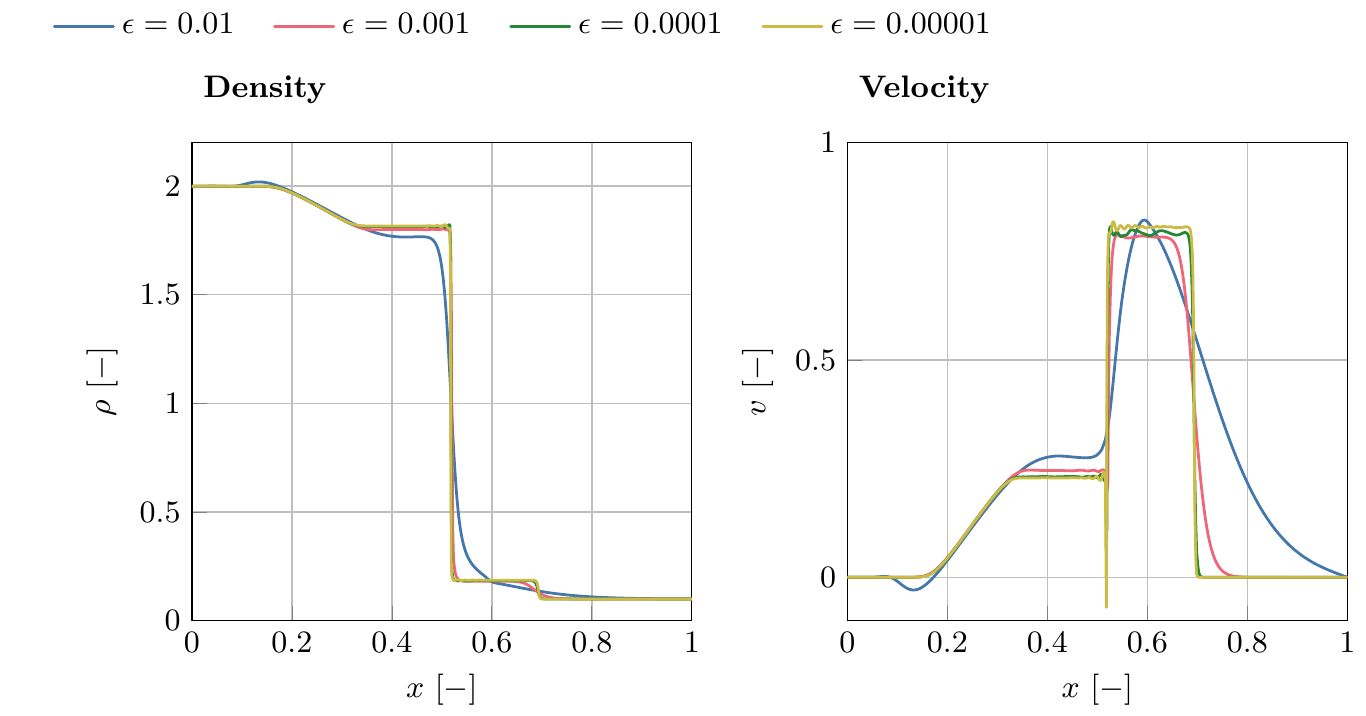}
   \caption{Density and velocity profiles of the multiphase shock tube problem at $t=0.11$ using \emph{relaxation model I}.
            }
   \label{fig:results_nsk_riemann_model1}
\end{figure}

To further discuss the overshoots, \emph{relaxation model I} was also used to simulate the Riemann problem.
The Lax-Friedrichs method was used as numerical flux function to avoid imaginary numbers of the speed of sound.
The results for density and velocity at $t=0.11$ are shown in \cref{fig:results_nsk_riemann_model1}.
The overshoot at the right moving shock wave vanishes but velocity oscillations appear around the phase boundary.
They increase in magnitude for smaller values for $\epsilonkorteweg$.
Furthermore, the plateau values of the inner states, left and right of the phase boundary, change.
In the case of a fully smeared phase interface, where $\epsilonkorteweg=0.01$, the simulation is identical to the one produced by \emph{relaxation model II}.
We assume that difference between both models stem from the different treatment of the non-conservative flux regarding model I and model II.
While model II provides a monotone pressure function, there is an additional contribution of the density gradient to the surface fluxes which was not considered using model I.
We note that the use of a path-conservative scheme did not resolve this issue.

\subsection{2D Test Cases}
\label{sec:results_validation_2d}

For all 2D test cases, the following parameters were kept the same.
The relaxation parameters were $\alphakorteweg=100$, $\betakorteweg=0.001$, the viscosity and the capillarity were $\epsilonkorteweg=0.01$ and $\gammakorteweg=1$, respectively.
The domain, $\Omega = (0,1)^2$, was discretized with $100^2$ elements and a polynomial degree of $N=3$.
The numerical flux function was the HLLC solver and time integration was implicit with $\cfl=100$ using the fourth-order ESDIRK with six stages, unless otherwise noted.

\paragraph{Merging and evaporating droplets}
Several test cases were conducted with two or more droplets that merge or evaporate to a single drop in equilibrium.
For this, the average density must be between the Maxwellian densities.
The initial conditions for $n$ droplets were
\begin{align}
\label{eq:results_nsk_2d_merge_initialcondition}
\rho(\x,t=0) & = \frac{\l( n \rho_{\liq}^{\sat} - \l(n-2\r) \rho_{\vap}^{\sat} \r)}{2} + \sum_{i=1}^{n} \frac{\rho_{\vap}^{\sat}-\rho_{\liq}^{\sat}}{2} \mathrm{tanh} \l( \frac{d_i-r_i}{2\sqrt{\gammakorteweg \epsilonkorteweg^2}} \r), \\
\v(\x,t=0) & = \l(0,0,0\r)\transpose
,
\end{align}
where $d_i=||\x-\x_i^0||$ is the Euclidean distance to the centre $\x_i^0$ of the $i$-th droplet and $r_i$ is its radius.

The first test case involved two droplets with radii $r_1=0.2$, $r_2=0.1$ at centres $\x^0_1=(0.4,0.5,0)\transpose$, $\x^0_2=(0.7,0.5,0)\transpose$, respectively.
In a second test case, four droplets were initialized at $\x^0_1=(0.3,0.5,0)\transpose$, $\x^0_2=(0.7,0.3,0)\transpose$,  $\x^0_3=(0.6,0.7,0)\transpose$,  $\x^0_4=(0.5,0.5,0)\transpose$
with radii $r_1=0.15$, $r_2=0.05$, $r_3=0.07$, $r_4=0.03$.
The third test case initially contained 101 droplets. Their positions and radii are listed in \cref{tbl:app_nsk_100drops}.
\begin{table}[!ht]
\caption{Initial data for 101 droplets}
\centering
\scriptsize
\begin{tabular}{l|c|c||l|c|c}
\toprule
ID & $r_i~[\si{-}]$ & $\x^0_i~[\si{-}]$ & ID & $r_i~[\si{-}]$ & $\x^0_i~[\si{-}]$ \\
\midrule
1  & 0.0124 & $( 0.54698,0.71747)\transpose$ & 52  & 0.0119 & $( 0.38085,0.61939)\transpose$ \\
2  & 0.014  & $( 0.40297,0.13343)\transpose$ & 53  & 0.0122 & $( 0.63458,0.61533)\transpose$ \\
3  & 0.015  & $( 0.10704,0.44579)\transpose$ & 54  & 0.0113 & $( 0.36323,0.12262)\transpose$ \\
4  & 0.0111 & $( 0.72417,0.50879)\transpose$ & 55  & 0.0126 & $( 0.40762,0.12379)\transpose$ \\
5  & 0.0126 & $( 0.61368,0.53049)\transpose$ & 56  & 0.0128 & $( 0.3687,0.28446 )\transpose$ \\
6  & 0.0118 & $( 0.78297,0.85972)\transpose$ & 57  & 0.0123 & $( 0.4684,0.73573 )\transpose$ \\
7  & 0.0118 & $( 0.56662,0.67772)\transpose$ & 58  & 0.019  & $( 0.50341,0.41131)\transpose$ \\
8  & 0.0128 & $( 0.81132,0.80584)\transpose$ & 59  & 0.0124 & $( 0.91054,0.82898)\transpose$ \\
9  & 0.0118 & $( 0.57678,0.53124)\transpose$ & 60  & 0.015  & $( 0.20643,0.93511)\transpose$ \\
10 & 0.011  & $( 0.94403,0.9559 )\transpose$ & 61  & 0.0116 & $( 0.3386,0.39907 )\transpose$ \\
11 & 0.0125 & $(0.87145,0.066677)\transpose$ & 62  & 0.0119 & $(0.57413,0.052211)\transpose$ \\
12 & 0.0119 & $( 0.5076,0.54152 )\transpose$ & 63  & 0.0122 & $( 0.48693,0.57119)\transpose$ \\
13 & 0.0115 & $( 0.78882,0.28166)\transpose$ & 64  & 0.0112 & $( 0.26222,0.74767)\transpose$ \\
14 & 0.019  & $( 0.47303,0.4809 )\transpose$ & 65  & 0.0122 & $( 0.57959,0.32024)\transpose$ \\
15 & 0.018  & $( 0.8288,0.68486 )\transpose$ & 66  & 0.0127 & $( 0.87833,0.49293)\transpose$ \\
16 & 0.0115 & $( 0.32248,0.20826)\transpose$ & 67  & 0.012  & $( 0.06095,0.22165)\transpose$ \\
17 & 0.017  & $( 0.97615,0.60816)\transpose$ & 68  & 0.015  & $( 0.44088,0.93927)\transpose$ \\
18 & 0.012  & $( 0.27821,0.32618)\transpose$ & 69  & 0.0126 & $(0.084258,0.48231)\transpose$ \\
19 & 0.016  & $(0.072831,0.88085)\transpose$ & 70  & 0.0113 & $( 0.56324,0.54   )\transpose$ \\
20 & 0.018  & $( 0.75122,0.13339)\transpose$ & 71  & 0.0113 & $( 0.53931,0.22106)\transpose$ \\
21 & 0.016  & $( 0.83119,0.10241)\transpose$ & 72  & 0.0129 & $(0.76806,0.095945)\transpose$ \\
22 & 0.0119 & $( 0.92234,0.95912)\transpose$ & 73  & 0.0123 & $(0.23309,0.060165)\transpose$ \\
23 & 0.0125 & $( 0.32702,0.1529 )\transpose$ & 74  & 0.0130 & $( 0.58736,0.81951)\transpose$ \\
24 & 0.0125 & $( 0.80407,0.15254)\transpose$ & 75  & 0.018  & $( 0.45897,0.77148)\transpose$ \\
25 & 0.0125 & $( 0.53825,0.15555)\transpose$ & 76  & 0.013  & $( 0.86098,0.1957 )\transpose$ \\
26 & 0.0122 & $(0.46329,0.089569)\transpose$ & 77  & 0.0112 & $( 0.66084,0.89512)\transpose$ \\
27 & 0.0126 & $( 0.82075,0.45442)\transpose$ & 78  & 0.0116 & $( 0.35388,0.6843 )\transpose$ \\
28 & 0.0124 & $( 0.95191,0.6689 )\transpose$ & 79  & 0.0118 & $( 0.34719,0.65685)\transpose$ \\
29 & 0.017  & $( 0.076273,0.8313)\transpose$ & 80  & 0.013  & $( 0.25372,0.99038)\transpose$ \\
30 & 0.013  & $( 0.70867,0.79024)\transpose$ & 81  & 0.0115 & $(0.95253,0.033692)\transpose$ \\
31 & 0.013  & $( 0.23493,0.71271)\transpose$ & 82  & 0.0113 & $( 0.2982,0.42425 )\transpose$ \\
32 & 0.012  & $( 0.3989,0.4726  )\transpose$ & 83  & 0.013  & $( 0.15841,0.48998)\transpose$ \\
33 & 0.017  & $( 0.26812,0.70859)\transpose$ & 84  & 0.018  & $( 0.3613,0.5835  )\transpose$ \\
34 & 0.018  & $( 0.83251,0.95806)\transpose$ & 85  & 0.012  & $( 0.74163,0.08327)\transpose$ \\
35 & 0.0111 & $( 0.99537,0.50578)\transpose$ & 86  & 0.0129 & $( 0.7059,0.66015 )\transpose$ \\
36 & 0.016  & $( 0.64975,0.30505)\transpose$ & 87  & 0.0121 & $(0.70089,0.052305)\transpose$ \\
37 & 0.0116 & $( 0.70395,0.78981)\transpose$ & 88  & 0.019  & $(0.006226,0.55683)\transpose$ \\
38 & 0.0129 & $( 0.9323,0.23639 )\transpose$ & 89  & 0.0116 & $( 0.37435,0.71203)\transpose$ \\
39 & 0.0112 & $( 0.68765,0.2343 )\transpose$ & 90  & 0.011  & $( 0.9015,0.48791 )\transpose$ \\
40 & 0.0118 & $( 0.56835,0.4647 )\transpose$ & 91  & 0.0127 & $( 0.31834,0.6176 )\transpose$ \\
41 & 0.0127 & $( 0.31834,0.6176 )\transpose$ & 92  & 0.0113 & $( 0.59708,0.21378)\transpose$ \\
42 & 0.0110 & $( 0.2978,0.64566 )\transpose$ & 93  & 0.0129 & $( 0.12501,0.38064)\transpose$ \\
43 & 0.0114 & $( 0.38836,0.10371)\transpose$ & 94  & 0.019  & $( 0.81769,0.37751)\transpose$ \\
44 & 0.013  & $( 0.98118,0.26286)\transpose$ & 95  & 0.0118 & $( 0.86199,0.24129)\transpose$ \\
45 & 0.015  & $(0.083821,0.62292)\transpose$ & 96  & 0.013  & $( 0.33771,0.52293)\transpose$ \\
46 & 0.0118 & $( 0.23613,0.41324)\transpose$ & 97  & 0.019  & $( 0.31781,0.21779)\transpose$ \\
47 & 0.0111 & $( 0.98445,0.85855)\transpose$ & 98  & 0.0122 & $( 0.54825,0.86101)\transpose$ \\
48 & 0.0126 & $( 0.74925,0.28394)\transpose$ & 99  & 0.0111 & $( 0.84185,0.61539)\transpose$ \\
49 & 0.0129 & $( 0.16689,0.77949)\transpose$ & 100 & 0.0129 & $( 0.9031,0.95485 )\transpose$ \\
50 & 0.017  & $( 0.10512,0.9196 )\transpose$ & 101 & 0.0124 & $( 0.74509,0.38482)\transpose$ \\
51 & 0.0119 & $( 0.72937,0.16264)\transpose$ & &    \\
\bottomrule
\end{tabular}
\label{tbl:app_nsk_100drops}
\end{table}

\begin{figure}[!ht]
   \centering
   \includegraphics[width=\linewidth]{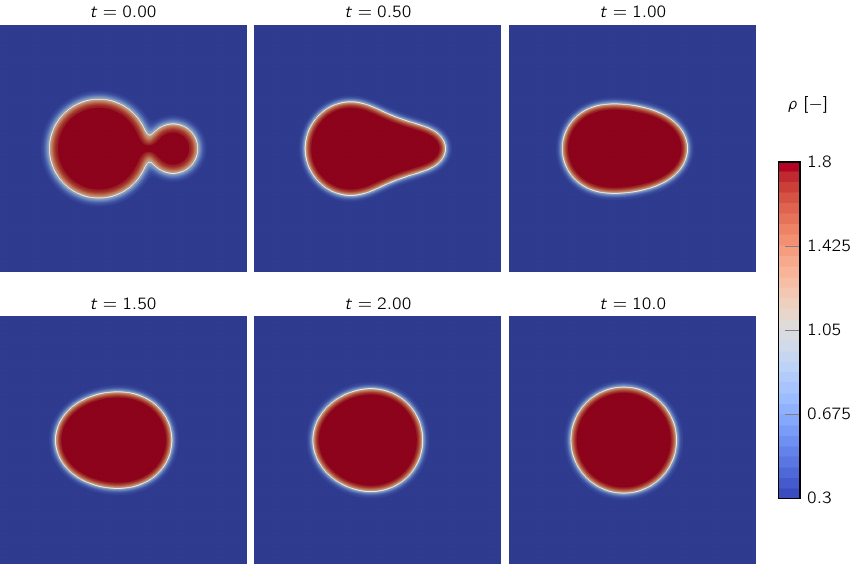}
   \caption{
             Density of two merging droplets at different time instances.
            }
   \label{fig:results_nsk_2d_drop_merge_2}
\end{figure}

\begin{figure}[!ht]
   \centering
   \includegraphics[width=\linewidth]{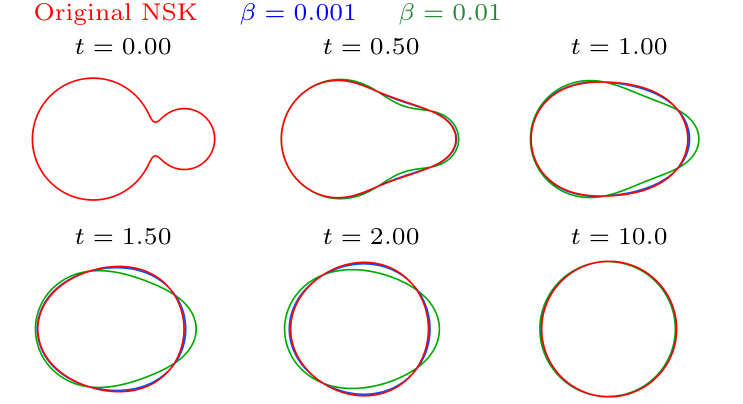}
   \caption{
             Contour of mean density, $\rho=1.0634$, of two merging droplets at different time instances.
             Comparison between relaxation model and original NSK equations.
            }
   \label{fig:results_nsk_2d_drop_merge_2_comparison}
\end{figure}

\Cref{fig:results_nsk_2d_drop_merge_2} shows the density of the two droplets at different time instances.
Although they were initially at rest, they move towards each other and eventually merge into a single, quiescent droplet.
This result comes close to the solution produced by the original NSK model.
The contour of the mean density, $\rho=1.0634$,  is presented in \cref{fig:results_nsk_2d_drop_merge_2_comparison} for the relaxation model and the NSK model.
In addition, a simulation with $\betakorteweg=0.01$ is presented.
In the end, all simulations reach the same equilibrium state, but their time evolution differ.
Similar to the unsteady 1D simulations discussed above, the choice of $\betakorteweg=0.001$ produces the best approximation of the original NSK model.
For the remainder of the simulations shown in this work, $\betakorteweg=0.001$ was used.

\begin{figure}[!ht]
   \centering
   \includegraphics[width=\linewidth]{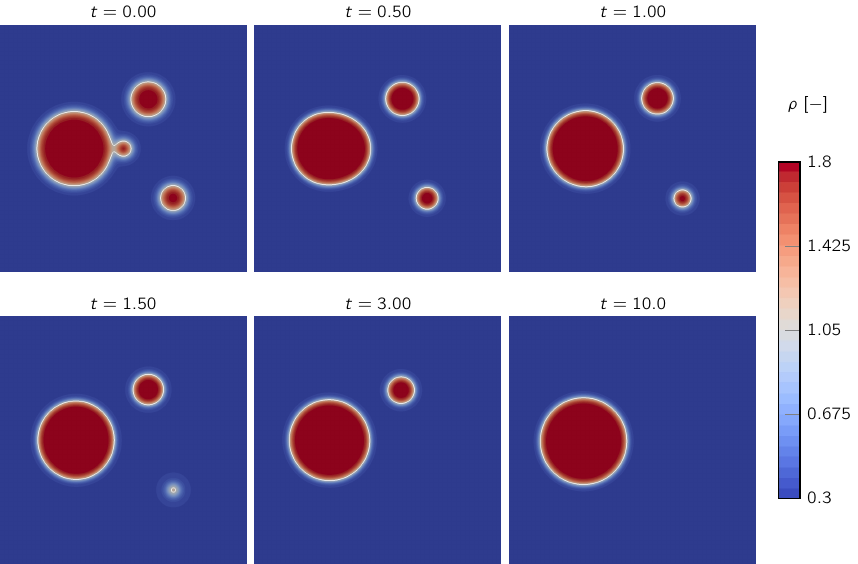}
   \caption{ Density of four merging and evaporating droplets at different time instances.
            }
   \label{fig:results_nsk_2d_drop_merge_4}
\end{figure}

In case of four initial droplets, shown in \cref{fig:results_nsk_2d_drop_merge_4}, the same behaviour was observed for the largest and smallest droplets. 
Initially they were positioned such that they almost touch which triggers coalescence.
The remaining two smaller droplets were positioned further away from their neighbours.
As time progresses they start to evaporate and the mass transfers to the largest droplet which in turn grows in size.
Eventually, the smaller droplets vanish and a single, large droplet remains.

\begin{figure}[!ht]
   \centering
   \includegraphics[width=\linewidth]{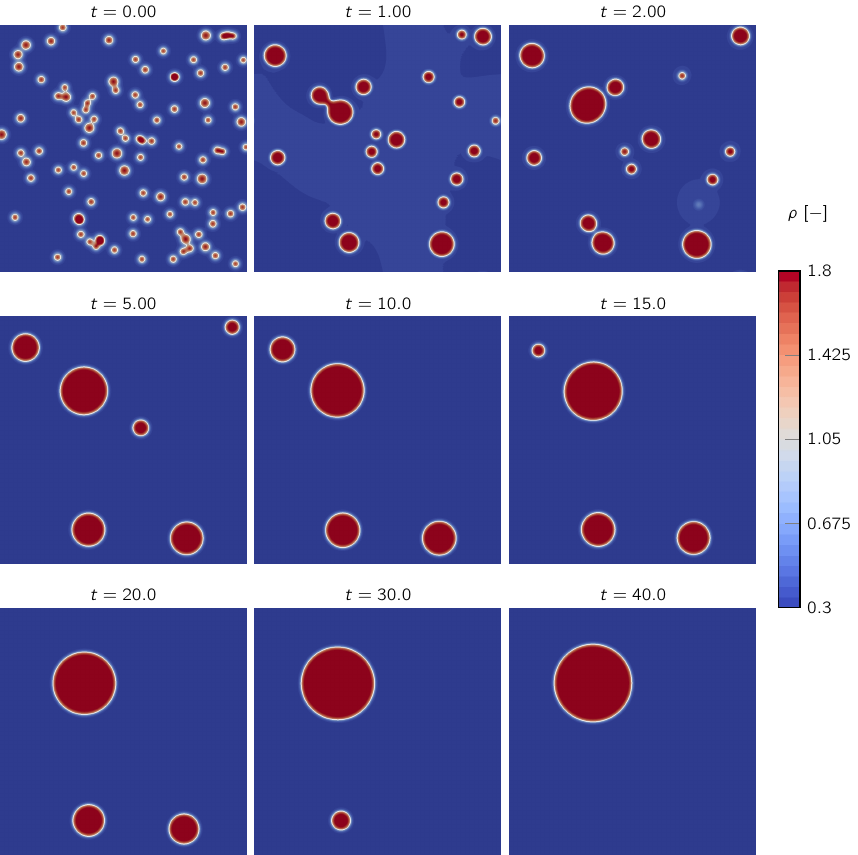}
   \caption{ Density of 101 merging and evaporating droplets at different time instances.
            }
   \label{fig:results_nsk_2d_many_drops}
\end{figure}

\Cref{fig:results_nsk_2d_many_drops} shows the case of 101 droplets as initial state.
Both coalescence and evaporation occurr depending on the distance of the droplets.
The smallest individual droplets evaporate and eventually droplet clusters merge into larger droplets.
These droplets are too far from each other to coalesce and they evaporate slowly.
The mass is eventually transferred to a finally remaining droplet which is in equilibrium with the surrounding vapour.

\begin{figure}[!ht]
   \centering
   \includegraphics[width=\linewidth]{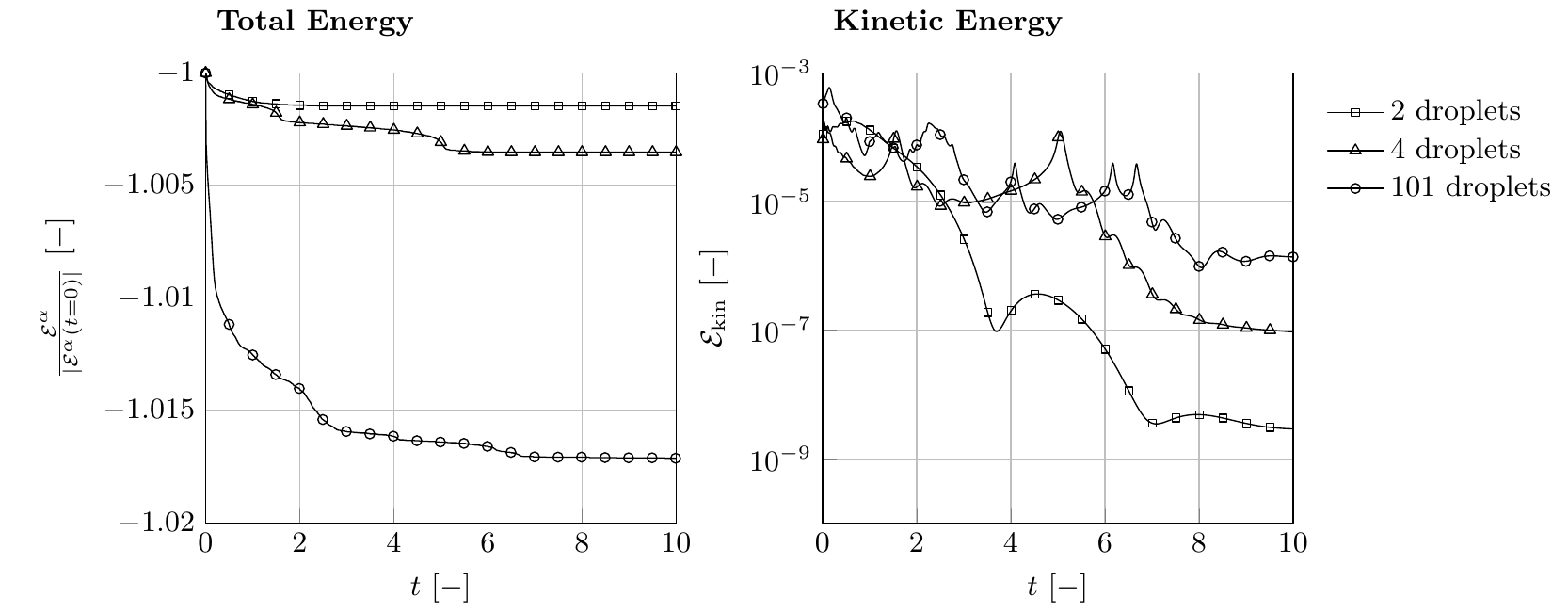}
   \caption{Energy decay of 2, 4, and 101 merging and evaporating droplets.
            }
   \label{fig:results_nsk_2d_merging_energydecay}
\end{figure}

\Cref{fig:results_nsk_2d_merging_energydecay} shows the decay of total and kinetic energy of the merging droplet cases.
For each number of droplets, both energies found a minimum.
In the kinetic energy, each disappearance of a droplet is marked by a local minimum of the energy, followed by a minor increase in flow movement.

\subsection{3D Test Cases: Head On Collisions\label{sec_3D}}

Binary head on collisions of droplets were simulated in 3D.
Four different cases were simulated with varying collision Weber and Reynolds numbers.
The simulation setup was adapted from Gelissen \etal\cite{gelissen_simulations_2018}.
They defined numerical test cases based on collision Weber and Reynolds numbers that lie in different collision regimes.
The present simulations were carried out in the vicinity of the critical point, hence, experimental comparisons are not available.
In contrast to Gelissen \etal\cite{gelissen_simulations_2018}, the present model is fully isothermal which implies infinite heat conduction.
Therefore, evaporation effects become stronger since temperature is fixed to a constant value.

\subsubsection{Simulation Setup}
The initial conditions were
\begin{align}
\label{eq:results_nsk_hoc_initialcondition}
\rho(\x,t=0) & = \rho_{\mathrm{vap}} + \frac{\rho_{\mathrm{liq}}-\rho_{\mathrm{vap}}}{2} \sum_{i=1}^2 \l(\mathrm{tanh} \l( \frac{d_i-r_i}{2\sqrt{\gammakorteweg \epsilonkorteweg^2}} \r)  \r), \\
v_1(\x,t=0) & = \begin{cases}
\frac{v_{\mathrm{rel}}}{4} \l( 1 - \mathrm{tanh} \l(\frac{d_1-r_{\mathrm{d}}}{2\sqrt{\gammakorteweg \epsilonkorteweg^2}} \r)  \r) \quad & \text{if} \quad x<0.5, \\
\frac{- v_{\mathrm{rel}}}{4} \l( 1 - \mathrm{tanh} \l(\frac{d_2-r_{\mathrm{d}}}{2\sqrt{\gammakorteweg \epsilonkorteweg^2}} \r)  \r) \quad & \text{if} \quad x\geq 0.5, \\
\end{cases}\\
v_2(\x,t=0) & = v_3(\x,t=0)= 0,
\end{align}
with droplet radii $r_1=r_2=0.1$ and distance $d_i = \parallel \x - \x^0_{i} \parallel$.
The initial positions of the droplets were $\x^0_{1}=(0.3,0.5,0.5)\transpose$ and $\x^0_{2}=(0.7,0.5,0.5)\transpose$.
The relaxation model parameters were $\alphakorteweg=100$ and $\betakorteweg=0.001$.
The capillarity coefficient was calculated from a fixed Weber number in all cases, $\We=6000=1/(\epsilonkorteweg^2 \gammakorteweg)$.
An additional non-dimensional number is the collision Weber number,
\begin{equation}
\label{eq:results_nsk_weber_collision}
\We_{\mathrm{coll}} = \frac{2 r \rho_{\mathrm{c}} v_{\mathrm{rel}}}{\sigma}
,
\end{equation}
where $\rho_{\mathrm{c}}=1$ is the critical density and $v_{\mathrm{rel}}$ is the relative velocity of the droplets.
For given values of $\We_{\mathrm{coll}}$ and $\mathrm{Re}$, the surface tension coefficient, $\sigma$, is calculated.
Together with the Young-Laplace law for spherical droplets, the initial pressures of the liquid and vapour phases were obtained similar to \cite{gelissen_simulations_2018} as,
\begin{equation}
\label{eq:results_nsk_hoc_initialpressures}
p_{\liq} = p^{\sat} + \frac{\rho_{\liq}^{\sat}}{\rho_{\liq}^{\sat} - \rho_{\vap}^{\sat}} \frac{2 \sigma}{r}
, \quad \text{and} \quad
p_{\vap} = p^{\sat} + \frac{\rho_{\vap}^{\sat}}{\rho_{\liq}^{\sat} - \rho_{\vap}^{\sat}} \frac{2 \sigma}{r}
.
\end{equation}
Based on these pressures, the initial densities of the phases, $\rho_{\liq}$ and $\rho_{\vap}$, were calculated for the temperature $T_{\mathrm{ref}}=0.85$. 

All relevant parameters for the simulations are given in \cref{tbl:results_nsk_hoc_initialdata}.
Cases HOC1, HOC2, and HOC3 use the same setup as in Gelissen \etal\cite{gelissen_simulations_2018}.
The fourth test case, HOC2A, is a modification of HOC2 with a slightly increased collision Weber number.
\begin{table}[ht]
\caption{Parameters for HOC cases. All numbers are non-dimensionalized.}
\centering
\begin{tabular}{l|c|c|c|c|c|c|c}
\toprule
 & $\mathrm{Re}$ & $\We_{\mathrm{coll}}$ & $\epsilonkorteweg$ & $\gammakorteweg$ & $v_{\mathrm{rel}}$ & $\rho_{\liq}$ & $\rho_{\vap}$ \\
\midrule
HOC1   & 300  & 75  & $\num{3.33d-3}$ & 15       & 1.5 & 1.8450 & 0.3496 \\
HOC2   & 1250 & 140 & $\num{8.00d-4}$ & 260.417 & 2.2 & 1.8502 & 0.3545 \\
HOC2A  & 1250 & 200 & $\num{8.00d-4}$ & 260.417 & 2.8 & 1.8553 & 0.3596 \\
HOC3   & 1000 & 520 & $\num{1.00d-3}$ & 166.667 & 4.0 & 1.8459 & 0.3504 \\
\bottomrule
\end{tabular}
\label{tbl:results_nsk_hoc_initialdata}
\end{table}
The computational domain, $\Omega = (0,1)^3$, was discretized by $64^3$ elements and a polynomial degree of $N=3$, resulting in $256^3$ DOFs.
The numerical flux function was the HLLC solver and time integration was implicit with $\cfl=100$ using the fourth-order ESDIRK method with six stages.
The simulations were conducted on the Cray XC40 (Hazel Hen) supercomputer at the High Performance Computing Center Stuttgart (HLRS) using 200 computational nodes (i.e. 4800 processors).

\subsubsection{Simulation Results}

The cases under investigation differ in the collision velocity and the collision Weber number.
Consequently, a different behaviour during time evolution was observed for each case.
The results are visualized by the iso-contour of the mean density, calculated from the initial densities, $\rho_{\mathrm{mean}}=(\rho_{\liq}+\rho_{\vap})/2$.

\begin{figure}[!ht]
   \centering
   \includegraphics[width=\linewidth]{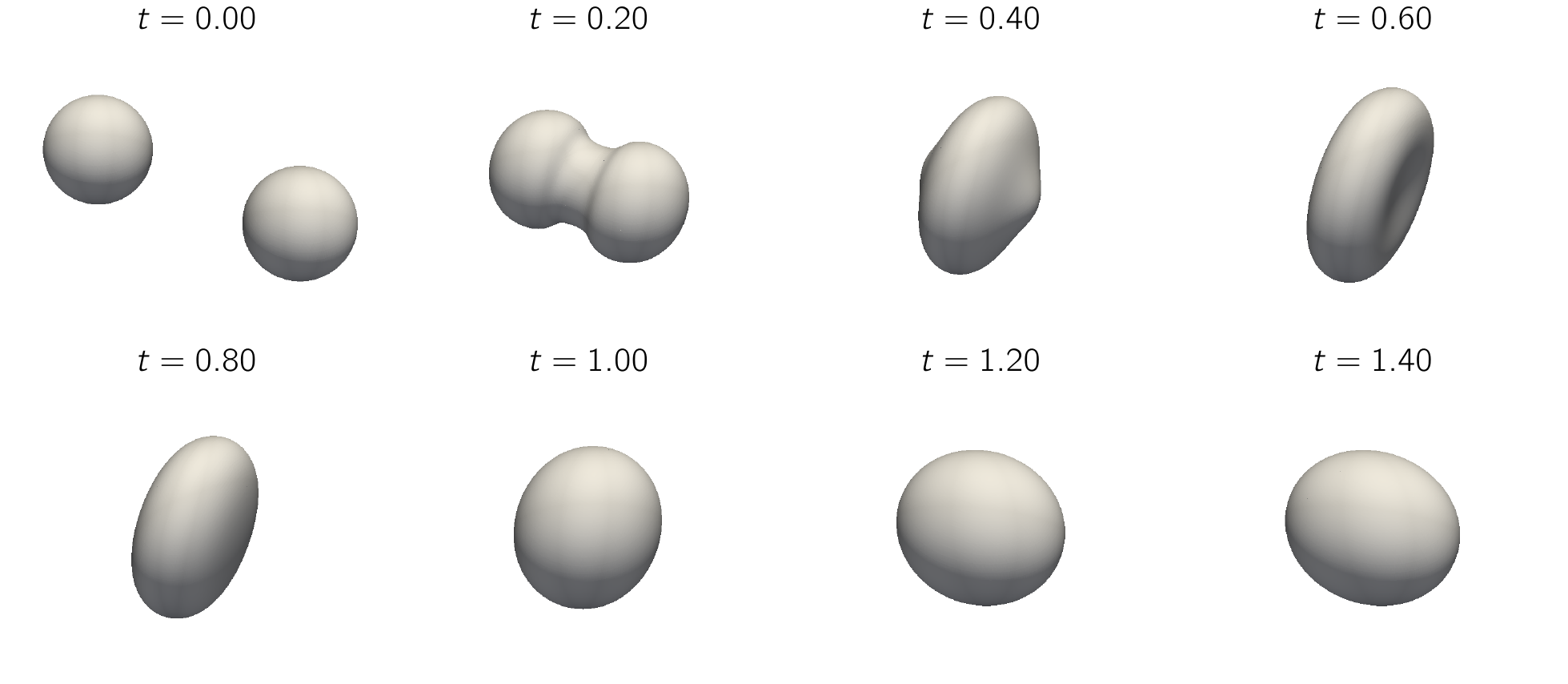}
   \caption{Mean density iso-contour for the HOC1 case at different time instances.
            }
   \label{fig:results_nsk_hoc1}
\end{figure}

\Cref{fig:results_nsk_hoc1} shows the time evolution of the first case with the smallest velocity and collision Weber number.
The two droplets move towards each other and at a critical distance, a liquid bridge forms which connects both droplets.
Then, coalescence occurs and the joined droplet takes the form of a flat disc.
The initial momentum is conserved and begins to interact with surface tension such that the disc starts to oscillate as a singular droplet.
This motion is damped due to viscous forces. 
In the end, the liquid phase is completely evaporated since the average density of the domain lies in a stable vapour phase.

\begin{figure}[!ht]
   \centering
   \includegraphics[width=\linewidth]{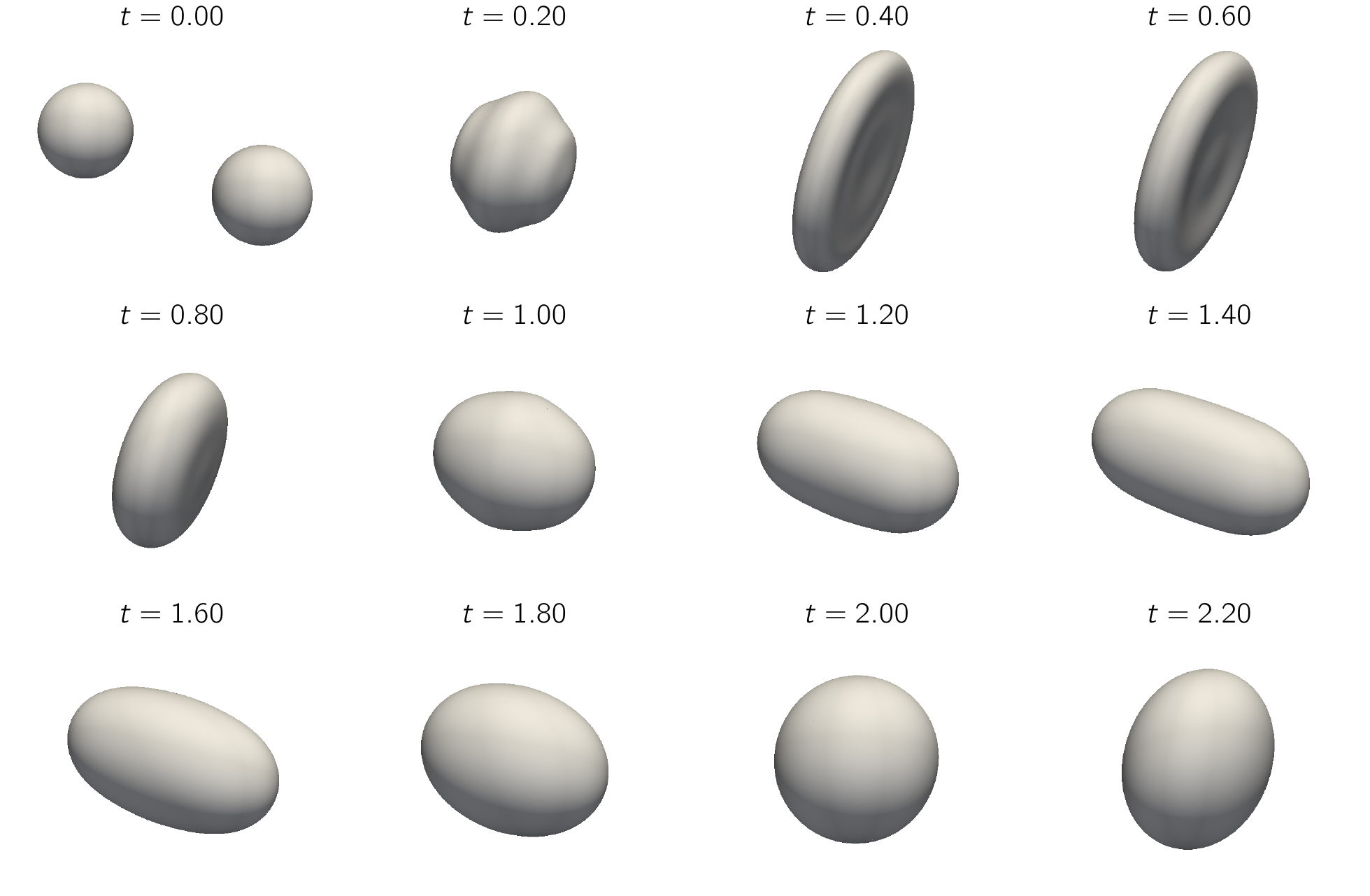}
   \caption{Mean density iso-contour for the HOC2 case at different time instances.
            }
   \label{fig:results_nsk_hoc2}
\end{figure}

\begin{figure}[!ht]
   \centering
   \includegraphics[width=\linewidth]{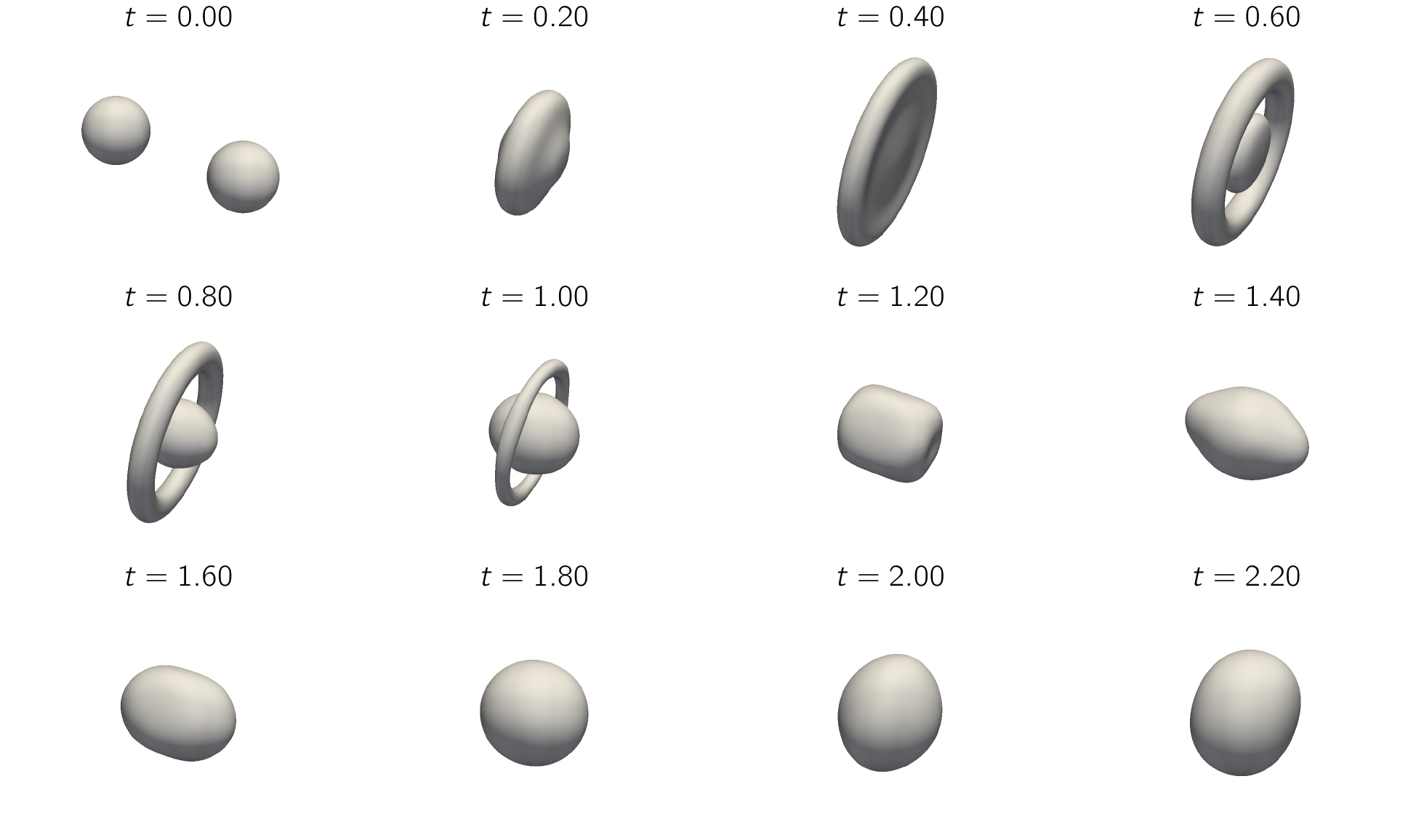}
   \caption{Mean density iso-contour for the HOC2A case at different time instances.
            }
   \label{fig:results_nsk_hoc2a}
\end{figure}

Increasing both impact velocity and Weber number amplifies the effects, as visualized in \cref{fig:results_nsk_hoc2} for the case HOC2.
The process evolves quicker and the flat disc obtains a larger radius.
A thin film is created in the centre of the disc which remains intact.
The surface tension forces cause the disc to return to a droplet like shape but the momentum is large enough that, during the first period of the oscillation, an elongated, round-ended cylinder shape forms.
Viscous damping and surface tension is strong enough to keep the droplet intact.
It remains oscillating until the liquid phase evaporates completely.
However, the results observed by Gelissen \etal\cite{gelissen_simulations_2018} are different.
For this collision Weber number, they reported reflexive separation, i.e. the elongated, cylindrical droplet split into two separate droplets.
A small increase in collision Weber number and impact velocity was simulated in case HOC2A.
The solution is visualized in \cref{fig:results_nsk_hoc2a}.
It shows that instead of reflexive separation a liquid ring is produced that evaporates quickly while a centre droplet remains oscillating.
Further increase of the collision Weber number does not produce any separation process.
It is assumed that neglecting temperature variations in the isothermal model increases damping effects, e.g. due to evaporation, such that the case of reflexive separation is suppressed.

\begin{figure}[!ht]
   \centering
   \includegraphics[width=\linewidth]{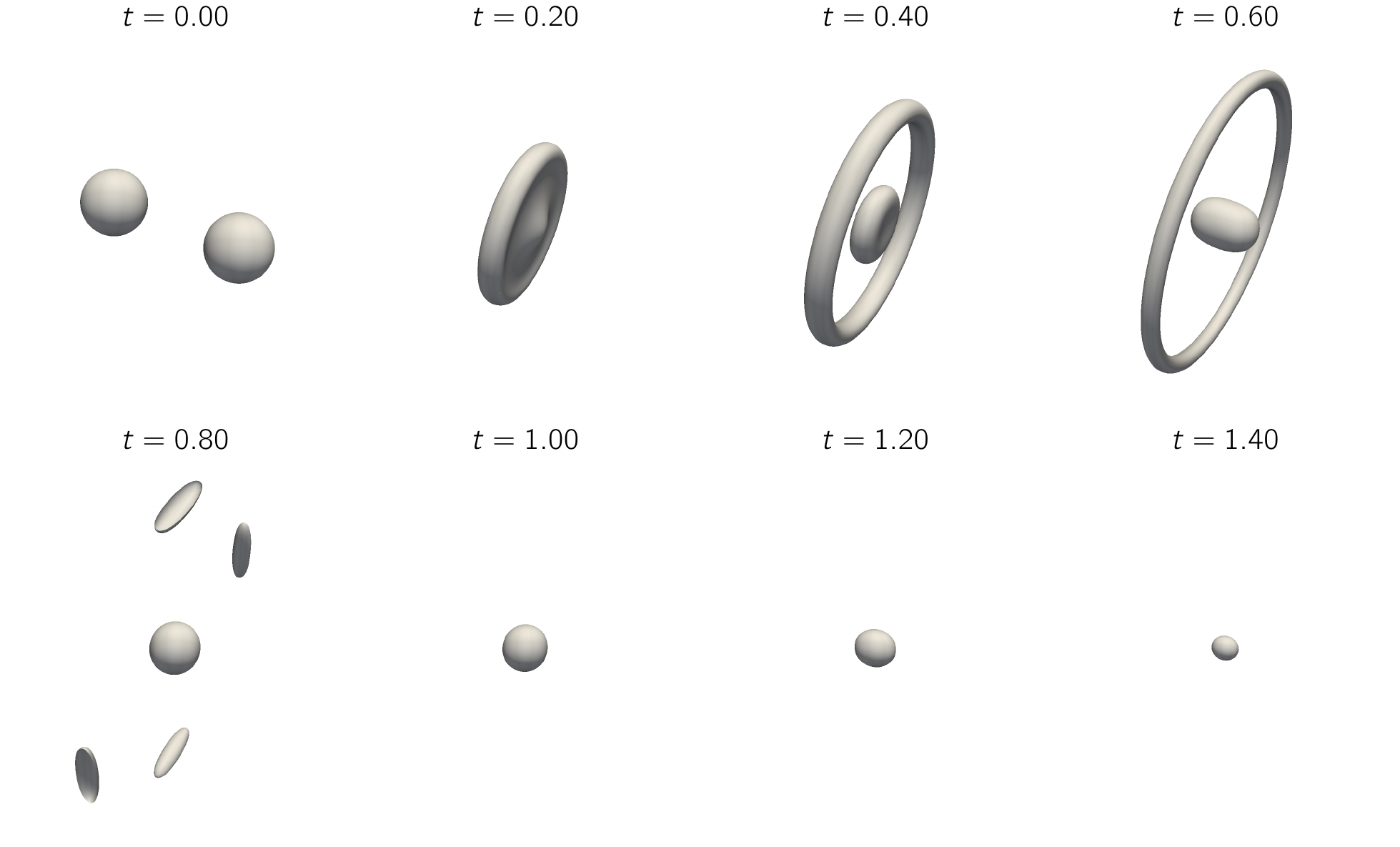}
   \caption{Mean density iso-contour for the HOC3 case at different time instances.
            }
   \label{fig:results_nsk_hoc3}
\end{figure}

An even higher velocity and collision Weber number is shown in \cref{fig:results_nsk_hoc3}.
The thin film of the flat disc becomes so thin that it breaks up and a liquid ring and a small droplet in its centre remains.
While the droplet in the domain centre starts to oscillate, the ring expands and increases its radius.
In proximity of the periodic domain boundary, the interaction with the periodic boundary condition triggers Plateau-Rayleigh instabilities \cite{gelissen_simulations_2018} and the ring contracts and forms four small droplets.
These are in non-equilibrium and immediately evaporate.
The droplet in the domain centre remains for a longer time period until it evaporates as well.
In contrast to the results presented here, Gelissen \etal\cite{gelissen_simulations_2018} reported for this collision Weber number that no centre droplet remains and that the ring splits into eight droplets which appear to remain stable.
The differences between the results of Gelissen \etal\cite{gelissen_simulations_2018} and the present work can be caused by several aspects.
First, the relaxation model is an approximation of the original NSK equations and therefore it may deal with interfacial instabilities differently during time advancement.
Furthermore, the present simulations were strictly isothermal which cause the liquid phase to evaporate and become a stable vapour phase in thermodynamic equilibrium.
Gelissen \etal\cite{gelissen_simulations_2018} on the other hand also discretized the energy equation and considered finite heat conduction.
Especially in the interfacial region, temperature peaks are reported which were not resolved in the present study.

\begin{figure}[!ht]
   \centering
   \includegraphics[width=\linewidth]{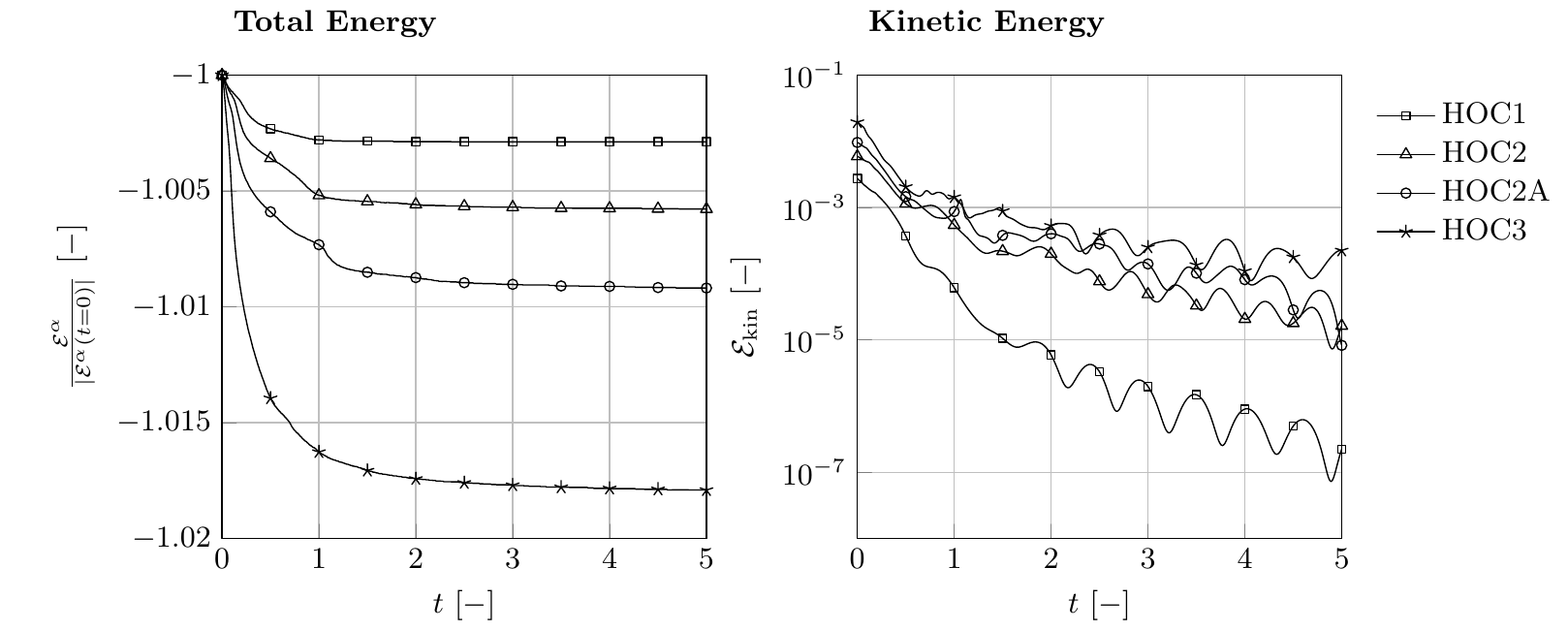}
   \caption{Decay of total and kinetic energy for the HOC1--HOC3 cases.
            }
   \label{fig:results_nsk_hoc_energydecay}
\end{figure}

For completeness, the decay of total and kinetic energy is shown in \cref{fig:results_nsk_hoc_energydecay} for all collision simulations.
A monotone decrease in total energy is observed while the kinetic energy decreases non-monotonously.

\section{Conclusion}
\label{sec:conclusion}

A lower-order parabolic relaxation model for the isothermal \nsk was investigated.
It adds a parabolic evolution equation for a relaxation variable, parametrized by the Korteweg parameter.
By use of a modified pressure function that takes into account the Korteweg parameter, a fully hyperbolic diffuse interface model can be constructed which allows for the straightforward implementation into the open source higher-order Discontinuous Galerkin solver \flexi.
Simulations in the Korteweg limit as well as in the  sharp interface  limit  become possible when utilizing a sub-cell shock capturing method.
The use of this  stabilization  approach applies also for extremely small values of the interfacial width parameter.
In the Korteweg limit, implicit time integration is preferred.

The model was validated against solutions of the original NSK model.
The test cases included static and travelling wave solutions in 1D and merging and evaporation of multiple droplets in 2D.
In the Korteweg limit, 3D simulations of binary head-on droplet collisions were simulated.
For low Weber numbers, the results are similar to findings from literature while for higher Weber number, the results differed since reflexive separation has not occurred.
This was attributed to the neglecting of temperature variations in the isothermal model.
Notably, in all cases, the free energy functional remains non-increasing in time rendering 
the discrete solutions to be thermodynamically admissible. 

Future work aims to extend the diffuse interface method to include heat transfer such that more realistic results can be obtained for simulations of, e.g., head-on collisions.
In the sharp interface limit, 2D and 3D problems, such as shock-droplet interactions, are of interest.
This requires a revisit of the FV shock capturing method in \flexi in terms of well-balanced and path-conservative schemes.
Alternatively, an adaptive mesh refinement method may be beneficial to locally refine the mesh only in the interfacial region while maintaining a coarse mesh where the solution remains near constant.

\section*{Acknowledgements}
The work was supported by the Deutsche Forschungsgemeinschaft (DFG, German Research Foundation) under Germany 's Excellence Strategy - EXC 2075 - 390740016 and through SFB-TRR 75 ``Droplet Dynamics under Extreme Ambient Conditions''. The simulations were performed on the national supercomputer Cray XC40 (Hazel Hen) at the High Performance Computing Center Stuttgart (HLRS).

  \bibliographystyle{elsarticle-num-names} 
  \bibliography{literature}


%
%
%
\end{document}

%% file: tikz_settings.tex
\usepackage{tikz}
\usepackage{tikz-3dplot}
\usepackage{xifthen}
\usetikzlibrary{spy,arrows}
\usetikzlibrary{arrows.meta}
\usetikzlibrary{shapes}
\usetikzlibrary{intersections}
\usetikzlibrary{positioning}
\usetikzlibrary{backgrounds}
\usepackage{pgfplots}
\usepackage{pgfplotstable}
\pgfplotsset{compat=1.13}
\usetikzlibrary{calc}
\tikzstyle{rounded}  = [rounded corners]     
\tikzstyle{function} = [rectangle, text centered, draw=black,minimum width=6.0cm,minimum height=1.0cm]
\tikzstyle{mpi}      = [anchor=south west, xshift=0.2cm] 

\usetikzlibrary{3d}

\usetikzlibrary{fillbetween}
\usetikzlibrary{decorations,decorations.text,backgrounds}



%% file: commands.tex
\renewcommand{\l}{\left}
\renewcommand{\r}{\right}

\newcommand{\gradient}{\nabla}
\newcommand{\gradientXI}{\nabla_\xi}

\renewcommand{\div}{\nabla \cdot}
\newcommand{\laplace}{\mathop{}\!\mathcal{4}}
\newcommand{\divXI}{\nabla_\xi \cdot}

\newcommand{\transpose}{^{\top}}

\newcommand{\half}{\frac{1}{2}}
\renewcommand{\d}{\operatorname{d}}

\newcommand{\dd}{\mbox{\,d}}

\newcommand{\intE}[1]{\int_E #1 \dd \xi}
\newcommand{\intpE}[1]{\int_{\partial E} #1 \dd S_\xi}

\newcommand{\fracp}[2]{\frac{\partial #1}{\partial #2}}

\newcommand{\oneover}[1]{\ensuremath{\frac{1}{#1}}}

\newcommand{\tensor}[1]{\uline{#1}}

\renewcommand{\v}{\mathbf{v}}
\newcommand{\U}{\mathbf{U}}

\newcommand{\Q}{\mathbf{Q}}
\newcommand{\source}{\mathbf{S}}
\newcommand{\sourceu}{\source(\U, \gradient \U)}
\newcommand{\I}{\tensor{\mathbf{I}}}

\newcommand{\x}{\mathbf{x}}

\newcommand{\F}{\mathcal{F}}

\newcommand{\Fu}{\F(\U, \gradient \U)}
\newcommand{\Fb}{\mathbf{F}}
\newcommand{\Fbu}{\Fb(\U, \gradient \U)}

\newcommand{\Fbc}{\Fb^{\mathrm{c}}}

\newcommand{\Fbv}{\Fb^{\mathrm{v}}}

\newcommand\hatU{\mathbf{\hat U}}

\newcommand{\cfl}{\mathrm{CFL}}

\newcommand{\jac}{\mathcal{J}}

\newcommand{\gammakorteweg}{\ensuremath{\gamma}}
\newcommand{\alphakorteweg}{\ensuremath{\alpha}}
\newcommand{\betakorteweg}{\ensuremath{\beta}}
\newcommand{\epsilonkorteweg}{\ensuremath{\epsilon}}
\newcommand{\ckorteweg}{\ensuremath{c}}
\newcommand{\stress}{\tensor{\boldsymbol{{\tau}}}}
\newcommand{\stresskorteweg}{\tensor{\boldsymbol{{\tau}}}_{\mathrm{K}}}

\newcommand{\avdw}{\ensuremath{a}}
\newcommand{\bvdw}{\ensuremath{b}}
\newcommand{\Rvdw}{\ensuremath{R}}

\newcommand{\We}{\ensuremath{\mathrm{We}}}

\providecommand{\nref}{}
\renewcommand{\nref}{\mathbf{n}_{E}}

\newcommand{\R}{\mathbb{R}} 
\newcommand{\N}{\mathbb{N}}

\newcommand{\liq}{\ensuremath{\mathrm{liq}}}
\newcommand{\vap}{\ensuremath{\mathrm{vap}}}
\newcommand{\sat}{\ensuremath{\mathrm{sat}}}

\newcommand{\nsk}{Navier-Stokes-Korteweg equations\xspace}

\newcommand{\flexi}{\textit{FLEXI}\xspace}

\newcommand{\etal}{et al.~}



%% file: 2019_nsk.bbl
\begin{thebibliography}{41}
\expandafter\ifx\csname natexlab\endcsname\relax\def\natexlab#1{#1}\fi
\providecommand{\url}[1]{\texttt{#1}}
\providecommand{\href}[2]{#2}
\providecommand{\path}[1]{#1}
\providecommand{\DOIprefix}{doi:}
\providecommand{\ArXivprefix}{arXiv:}
\providecommand{\URLprefix}{URL: }
\providecommand{\Pubmedprefix}{pmid:}
\providecommand{\doi}[1]{\href{http://dx.doi.org/#1}{\path{#1}}}
\providecommand{\Pubmed}[1]{\href{pmid:#1}{\path{#1}}}
\providecommand{\bibinfo}[2]{#2}
\ifx\xfnm\relax \def\xfnm[#1]{\unskip,\space#1}\fi
\bibitem[{Ishii and Hibiki(2011)}]{ishii_thermo-fluid_2011}
\bibinfo{author}{M.~Ishii}, \bibinfo{author}{T.~Hibiki},
  \bibinfo{title}{Thermo-{Fluid} {Dynamics} of {Two}-{Phase} {Flow}},
  \bibinfo{edition}{2} ed., \bibinfo{publisher}{Springer-Verlag},
  \bibinfo{address}{New York}, \bibinfo{year}{2011}.
\bibitem[{Anderson et~al.(1998)Anderson, McFadden, and
  Wheeler}]{anderson_diffuse-interface_1998}
\bibinfo{author}{D.~M. Anderson}, \bibinfo{author}{G.~B. McFadden},
  \bibinfo{author}{A.~A. Wheeler},
\newblock \bibinfo{title}{Diffuse-{Interface} {Methods} in {Fluid}
  {Mechanics}},
\newblock \bibinfo{journal}{Annual Review of Fluid Mechanics}
  \bibinfo{volume}{30} (\bibinfo{year}{1998}) \bibinfo{pages}{139--165}.
  \DOIprefix\doi{10.1146/annurev.fluid.30.1.139}.
\bibitem[{Van~der Waals(1894)}]{van_der_waals_thermodynamische_1894}
\bibinfo{author}{J.~Van~der Waals},
\newblock \bibinfo{title}{Thermodynamische {Theorie} der {Kapillarität} unter
  voraussetzung stetiger {Dichteänderung}},
\newblock \bibinfo{journal}{Zeitschrift für Physikalische Chemie}
  \bibinfo{volume}{13} (\bibinfo{year}{1894}) \bibinfo{pages}{657--725}.
\bibitem[{Korteweg(1901)}]{korteweg_sur_1901}
\bibinfo{author}{D.~J. Korteweg},
\newblock \bibinfo{title}{Sur la forme que prennent les équations du
  mouvements des fluides si l'on tient compte des forces capillaires causées
  par des variations de densité considérables mais connues et sur la théorie
  de la capillarité dans l'hypothèse d'une variation continue de la
  densité},
\newblock \bibinfo{journal}{Archives Néerlandaises des Sciences exactes et
  naturelles} \bibinfo{volume}{6} (\bibinfo{year}{1901})
  \bibinfo{pages}{1--24}.
\bibitem[{Dunn and Serrin(1986)}]{dunn_thermomechanics_1986}
\bibinfo{author}{J.~E. Dunn}, \bibinfo{author}{J.~Serrin},
\newblock \bibinfo{title}{On the {Thermomechanics} of {Interstitial}
  {Working}},
\newblock in: \bibinfo{booktitle}{The {Breadth} and {Depth} of {Continuum}
  {Mechanics}}, \bibinfo{publisher}{Springer Berlin Heidelberg},
  \bibinfo{address}{Berlin, Heidelberg}, \bibinfo{year}{1986}, pp.
  \bibinfo{pages}{705--743}.
\bibitem[{Bresch et~al.(2003)Bresch, Desjardins, and
  Lin}]{bresch_compressible_2003}
\bibinfo{author}{D.~Bresch}, \bibinfo{author}{B.~Desjardins},
  \bibinfo{author}{C.-K. Lin},
\newblock \bibinfo{title}{On {Some} {Compressible} {Fluid} {Models}:
  {Korteweg}, {Lubrication}, and {Shallow} {Water} {Systems}},
\newblock \bibinfo{journal}{Communications in Partial Differential Equations}
  \bibinfo{volume}{28} (\bibinfo{year}{2003}) \bibinfo{pages}{843--868}.
  \DOIprefix\doi{10.1081/PDE-120020499}.
\bibitem[{Hattori and Li(1994)}]{hattori_solutions_1994}
\bibinfo{author}{H.~Hattori}, \bibinfo{author}{D.~Li},
\newblock \bibinfo{title}{Solutions for {Two}-{Dimensional} {System} for
  {Materials} of {Korteweg} {Type}},
\newblock \bibinfo{journal}{SIAM Journal on Mathematical Analysis}
  \bibinfo{volume}{25} (\bibinfo{year}{1994}) \bibinfo{pages}{85--98}.
  \DOIprefix\doi{10.1137/S003614109223413X}.
\bibitem[{Kotschote(2008)}]{kotschote_strong_2008}
\bibinfo{author}{M.~Kotschote},
\newblock \bibinfo{title}{Strong solutions for a compressible fluid model of
  {Korteweg} type},
\newblock \bibinfo{journal}{Annales de l'Institut Henri Poincare (C) Non Linear
  Analysis} \bibinfo{volume}{25} (\bibinfo{year}{2008}) \bibinfo{pages}{679 --
  696}. \DOIprefix\doi{10.1016/j.anihpc.2007.03.005}.
\bibitem[{Rohde(2005)}]{rohde_local_2005}
\bibinfo{author}{C.~Rohde},
\newblock \bibinfo{title}{On local and non-local {Navier}-{Stokes}-{Korteweg}
  systems for liquid-vapour phase transitions},
\newblock \bibinfo{journal}{ZAMM - Journal of Applied Mathematics and Mechanics
  / Zeitschrift für Angewandte Mathematik und Mechanik} \bibinfo{volume}{85}
  (\bibinfo{year}{2005}) \bibinfo{pages}{839--857}.
  \DOIprefix\doi{10.1002/zamm.200410211}.
\bibitem[{Braack and Prohl(2013)}]{braack_stable_2013}
\bibinfo{author}{M.~Braack}, \bibinfo{author}{A.~Prohl},
\newblock \bibinfo{title}{Stable discretization of a diffuse interface model
  for liquid-vapor flows with surface tension},
\newblock \bibinfo{journal}{ESAIM: Mathematical Modelling and Numerical
  Analysis} \bibinfo{volume}{47} (\bibinfo{year}{2013})
  \bibinfo{pages}{401--420}. \DOIprefix\doi{10.1051/m2an/2012032}.
\bibitem[{Coquel et~al.(2005)Coquel, Diehl, Merkle, and
  Rohde}]{coquel_sharp_2005}
\bibinfo{author}{F.~Coquel}, \bibinfo{author}{D.~Diehl},
  \bibinfo{author}{C.~Merkle}, \bibinfo{author}{C.~Rohde},
\newblock \bibinfo{title}{Sharp and diffuse interface methods for phase
  transition problems in liquid-vapour flows},
\newblock \bibinfo{journal}{Numerical methods for hyperbolic and kinetic
  problems} \bibinfo{volume}{7} (\bibinfo{year}{2005})
  \bibinfo{pages}{239--270}.
\bibitem[{Diehl(2007)}]{diehl_higher_2007}
\bibinfo{author}{D.~Diehl}, \bibinfo{title}{Higher {Order} {Schemes} for
  {Simulation} of {Compressible} {Liquid}-{Vapour} {Flows} with {Phase}
  {Change}}, \bibinfo{type}{{PhD} thesis}, Albert-Ludwigs-Universität,
  \bibinfo{address}{Freiburg}, \bibinfo{year}{2007}.
\bibitem[{Diehl et~al.(2016)Diehl, Kremser, Kröner, and
  Rohde}]{diehl_numerical_2016}
\bibinfo{author}{D.~Diehl}, \bibinfo{author}{J.~Kremser},
  \bibinfo{author}{D.~Kröner}, \bibinfo{author}{C.~Rohde},
\newblock \bibinfo{title}{Numerical solution of
  {Navier}–{Stokes}–{Korteweg} systems by {Local} {Discontinuous}
  {Galerkin} methods in multiple space dimensions},
\newblock \bibinfo{journal}{Applied Mathematics and Computation}
  \bibinfo{volume}{272} (\bibinfo{year}{2016}) \bibinfo{pages}{309 -- 335}.
  \DOIprefix\doi{10.1016/j.amc.2015.09.080}.
\bibitem[{Gelissen et~al.(2018)Gelissen, Geld, Kuipers, and
  Kuerten}]{gelissen_simulations_2018}
\bibinfo{author}{E.~J. Gelissen}, \bibinfo{author}{C.~W. M. v.~d. Geld},
  \bibinfo{author}{J.~A.~M. Kuipers}, \bibinfo{author}{J.~G.~M. Kuerten},
\newblock \bibinfo{title}{Simulations of droplet collisions with a {Diffuse}
  {Interface} {Model} near the critical point},
\newblock \bibinfo{journal}{International Journal of Multiphase Flow}
  \bibinfo{volume}{107} (\bibinfo{year}{2018}) \bibinfo{pages}{208 -- 220}.
  \DOIprefix\doi{10.1016/j.ijmultiphaseflow.2018.06.001}.
\bibitem[{Giesselmann et~al.(2014)Giesselmann, Makridakis, and
  Pryer}]{giesselmann_energy_2014}
\bibinfo{author}{J.~Giesselmann}, \bibinfo{author}{C.~Makridakis},
  \bibinfo{author}{T.~Pryer},
\newblock \bibinfo{title}{Energy consistent discontinuous {Galerkin} methods
  for the {Navier}–{Stokes}–{Korteweg} system},
\newblock \bibinfo{journal}{Mathematics of Computation} \bibinfo{volume}{83}
  (\bibinfo{year}{2014}) \bibinfo{pages}{2071--2099}.
\bibitem[{Gomez et~al.(2010)Gomez, Hughes, Nogueira, and
  Calo}]{gomez_isogeometric_2010}
\bibinfo{author}{H.~Gomez}, \bibinfo{author}{T.~J.~R. Hughes},
  \bibinfo{author}{X.~Nogueira}, \bibinfo{author}{V.~M. Calo},
\newblock \bibinfo{title}{Isogeometric analysis of the isothermal
  {Navier}–{Stokes}–{Korteweg} equations},
\newblock \bibinfo{journal}{Computer Methods in Applied Mechanics and
  Engineering} \bibinfo{volume}{199} (\bibinfo{year}{2010})
  \bibinfo{pages}{1828 -- 1840}. \DOIprefix\doi{10.1016/j.cma.2010.02.010}.
\bibitem[{Haink and Rohde(2008)}]{haink_local_2008}
\bibinfo{author}{J.~Haink}, \bibinfo{author}{C.~Rohde},
\newblock \bibinfo{title}{Local discontinuous-{Galerkin} schemes for model
  problems in phase transition theory},
\newblock \bibinfo{journal}{Commun. Comput. Phys} \bibinfo{volume}{4}
  (\bibinfo{year}{2008}) \bibinfo{pages}{860--893}.
\bibitem[{Jamet et~al.(2001)Jamet, Lebaigue, Coutris, and
  Delhaye}]{jamet_second_2001}
\bibinfo{author}{D.~Jamet}, \bibinfo{author}{O.~Lebaigue},
  \bibinfo{author}{N.~Coutris}, \bibinfo{author}{J.~M. Delhaye},
\newblock \bibinfo{title}{The {Second} {Gradient} {Method} for the {Direct}
  {Numerical} {Simulation} of {Liquid}–{Vapor} {Flows} with {Phase}
  {Change}},
\newblock \bibinfo{journal}{Journal of Computational Physics}
  \bibinfo{volume}{169} (\bibinfo{year}{2001}) \bibinfo{pages}{624 -- 651}.
  \DOIprefix\doi{10.1006/jcph.2000.6692}.
\bibitem[{Martínez et~al.(2019)Martínez, Ramírez, Nogueira, Khelladi, and
  Navarrina}]{martinez_high-order_2019}
\bibinfo{author}{A.~Martínez}, \bibinfo{author}{L.~Ramírez},
  \bibinfo{author}{X.~Nogueira}, \bibinfo{author}{S.~Khelladi},
  \bibinfo{author}{F.~Navarrina},
\newblock \bibinfo{title}{A high-order finite volume method with improved
  isotherms reconstruction for the computation of multiphase flows using the
  {Navier}-{Stokes}-{Korteweg} equations},
\newblock \bibinfo{journal}{Computers \& Mathematics with Applications}
  (\bibinfo{year}{2019}). \DOIprefix\doi{10.1016/j.camwa.2019.07.021}.
\bibitem[{Tian et~al.(2015)Tian, Xu, Kuerten, and Vegt}]{tian_local_2015}
\bibinfo{author}{L.~Tian}, \bibinfo{author}{Y.~Xu}, \bibinfo{author}{J.~G.~M.
  Kuerten}, \bibinfo{author}{J.~J. W. v.~d. Vegt},
\newblock \bibinfo{title}{A local discontinuous {Galerkin} method for the
  (non)-isothermal {Navier}–{Stokes}–{Korteweg} equations},
\newblock \bibinfo{journal}{Journal of Computational Physics}
  \bibinfo{volume}{295} (\bibinfo{year}{2015}) \bibinfo{pages}{685 -- 714}.
  \DOIprefix\doi{10.1016/j.jcp.2015.04.025}.
\bibitem[{Rohde(2010)}]{rohde_local_2010}
\bibinfo{author}{C.~Rohde},
\newblock \bibinfo{title}{A local and low-order {Navier}-{Stokes}-{Korteweg}
  system},
\newblock in: \bibinfo{booktitle}{Nonlinear {Partial} {Differential}
  {Equations} and {Hyperbolic} {Wave} {Phenomena}}, volume
  \bibinfo{volume}{526}, \bibinfo{publisher}{American Mathematical Society},
  \bibinfo{address}{Providence, RI}, \bibinfo{year}{2010}, pp.
  \bibinfo{pages}{315--337}.
\bibitem[{Neusser et~al.(2015)Neusser, Rohde, and
  Schleper}]{neusser_relaxation_2015}
\bibinfo{author}{J.~Neusser}, \bibinfo{author}{C.~Rohde},
  \bibinfo{author}{V.~Schleper},
\newblock \bibinfo{title}{Relaxation of the {Navier}-{Stokes}-{Korteweg}
  equations for compressible two-phase flow with phase transition},
\newblock \bibinfo{journal}{International Journal for Numerical Methods in
  Fluids} \bibinfo{volume}{79} (\bibinfo{year}{2015})
  \bibinfo{pages}{615--639}. \DOIprefix\doi{10.1002/fld.4065}.
\bibitem[{Corli et~al.(2014)Corli, Rohde, and Schleper}]{corli_parabolic_2014}
\bibinfo{author}{A.~Corli}, \bibinfo{author}{C.~Rohde},
  \bibinfo{author}{V.~Schleper},
\newblock \bibinfo{title}{Parabolic approximations of diffusive–dispersive
  equations},
\newblock \bibinfo{journal}{Journal of Mathematical Analysis and Applications}
  \bibinfo{volume}{414} (\bibinfo{year}{2014}) \bibinfo{pages}{773 -- 798}.
  \DOIprefix\doi{10.1016/j.jmaa.2014.01.049}.
\bibitem[{Rohde(2018)}]{rohde_fully_2018}
\bibinfo{author}{C.~Rohde},
\newblock \bibinfo{title}{Fully {Resolved} {Compressible} {Two}-{Phase} {Flow}:
  {Modelling}, {Analytical} and {Numerical} {Issues}},
\newblock in: \bibinfo{editor}{M.~Bulicek}, \bibinfo{editor}{E.~Feireisl},
  \bibinfo{editor}{M.~Pokorny} (Eds.), \bibinfo{booktitle}{New {Trends} and
  {Results} in {Mathematical} {Description} of {Fluid} {Flows}},
  \bibinfo{publisher}{Springer International Publishing},
  \bibinfo{address}{Cham}, \bibinfo{year}{2018}, pp. \bibinfo{pages}{115--181}.
  \DOIprefix\doi{10.1007/978-3-319-94343-5_4}.
\bibitem[{Hindenlang et~al.(2012)Hindenlang, Gassner, Altmann, Beck,
  Staudenmaier, and Munz}]{hindenlang_explicit_2012}
\bibinfo{author}{F.~Hindenlang}, \bibinfo{author}{G.~J. Gassner},
  \bibinfo{author}{C.~Altmann}, \bibinfo{author}{A.~Beck},
  \bibinfo{author}{M.~Staudenmaier}, \bibinfo{author}{C.-D. Munz},
\newblock \bibinfo{title}{Explicit discontinuous {Galerkin} methods for
  unsteady problems},
\newblock \bibinfo{journal}{Computers \& Fluids} \bibinfo{volume}{61}
  (\bibinfo{year}{2012}) \bibinfo{pages}{86--93}.
  \DOIprefix\doi{10.1016/j.compfluid.2012.03.006}.
\bibitem[{Sonntag and Munz(2014)}]{sonntag_shock_2014}
\bibinfo{author}{M.~Sonntag}, \bibinfo{author}{C.-D. Munz},
\newblock \bibinfo{title}{Shock {Capturing} for {Discontinuous} {Galerkin}
  {Methods} using {Finite} {Volume} {Subcells}},
\newblock in: \bibinfo{editor}{J.~Fuhrmann}, \bibinfo{editor}{M.~Ohlberger},
  \bibinfo{editor}{C.~Rohde} (Eds.), \bibinfo{booktitle}{Finite {Volumes} for
  {Complex} {Applications} {VII}-{Elliptic}, {Parabolic} and {Hyperbolic}
  {Problems}}, \bibinfo{publisher}{Springer International Publishing},
  \bibinfo{year}{2014}, pp. \bibinfo{pages}{945--953}.
  \DOIprefix\doi{10.1007/978-3-319-05591-6_96}.
\bibitem[{Sonntag and Munz(2017)}]{sonntag_efficient_2017}
\bibinfo{author}{M.~Sonntag}, \bibinfo{author}{C.-D. Munz},
\newblock \bibinfo{title}{Efficient {Parallelization} of a {Shock} {Capturing}
  for {Discontinuous} {Galerkin} {Methods} using {Finite} {Volume}
  {Sub}-cells},
\newblock \bibinfo{journal}{Journal of Scientific Computing}
  \bibinfo{volume}{70} (\bibinfo{year}{2017}) \bibinfo{pages}{1262--1289}.
  \DOIprefix\doi{10.1007/s10915-016-0287-5}.
\bibitem[{Giesselmann et~al.(2014)Giesselmann, Makridakis, and
  Pryer}]{GiesselmannMakridakis}
\bibinfo{author}{J.~Giesselmann}, \bibinfo{author}{C.~Makridakis},
  \bibinfo{author}{T.~Pryer},
\newblock \bibinfo{title}{Energy consistent discontinuous {G}alerkin methods
  for the {N}avier-{S}tokes-{K}orteweg system},
\newblock \bibinfo{journal}{Math. Comp.} \bibinfo{volume}{83}
  (\bibinfo{year}{2014}) \bibinfo{pages}{2071--2099}.
  \DOIprefix\doi{10.1090/S0025-5718-2014-02792-0}.
\bibitem[{Carr et~al.(1984)Carr, Gurtin, and Slemrod}]{carr_structured_1984}
\bibinfo{author}{J.~Carr}, \bibinfo{author}{M.~E. Gurtin},
  \bibinfo{author}{M.~Slemrod},
\newblock \bibinfo{title}{Structured phase transitions on a finite interval},
\newblock \bibinfo{journal}{Archive for rational mechanics and analysis}
  \bibinfo{volume}{86} (\bibinfo{year}{1984}) \bibinfo{pages}{317--351}.
  \DOIprefix\doi{10.1007/BF00280031}.
\bibitem[{Dreyer et~al.(2012)Dreyer, Giesselmann, Kraus, and
  Rohde}]{dreyer_asymptotic_2012}
\bibinfo{author}{W.~Dreyer}, \bibinfo{author}{J.~Giesselmann},
  \bibinfo{author}{C.~Kraus}, \bibinfo{author}{C.~Rohde},
\newblock \bibinfo{title}{Asymptotic analysis for {Korteweg} models},
\newblock \bibinfo{journal}{Interfaces and Free Boundaries}
  \bibinfo{volume}{14} (\bibinfo{year}{2012}) \bibinfo{pages}{105--143}.
  \DOIprefix\doi{10.4171/IFB/275}.
\bibitem[{Kopriva(2006)}]{kopriva_metric_2006}
\bibinfo{author}{D.~A. Kopriva},
\newblock \bibinfo{title}{Metric {Identities} and the {Discontinuous}
  {Spectral} {Element} {Method} on {Curvilinear} {Meshes}},
\newblock \bibinfo{journal}{Journal of Scientific Computing}
  \bibinfo{volume}{26} (\bibinfo{year}{2006}) \bibinfo{pages}{301--327}.
  \DOIprefix\doi{10.1007/s10915-005-9070-8}.
\bibitem[{Kopriva(2009)}]{kopriva_implementing_2009}
\bibinfo{author}{D.~A. Kopriva}, \bibinfo{title}{Implementing {Spectral}
  {Methods} for {Partial} {Differential} {Equations}},
  \bibinfo{publisher}{Springer}, \bibinfo{year}{2009}.
  \DOIprefix\doi{10.1007/978-90-481-2261-5_8}.
\bibitem[{Toro(2009)}]{toro_riemann_2009}
\bibinfo{author}{E.~F. Toro}, \bibinfo{title}{Riemann solvers and numerical
  methods for fluid dynamics: a practical introduction}, \bibinfo{edition}{2}
  ed., \bibinfo{publisher}{Springer Science \& Business Media},
  \bibinfo{address}{Berlin, Heidelberg}, \bibinfo{year}{2009}.
\bibitem[{Bassi and Rebay(1997)}]{bassi_high-order_1997}
\bibinfo{author}{F.~Bassi}, \bibinfo{author}{S.~Rebay},
\newblock \bibinfo{title}{A {High}-{Order} {Accurate} {Discontinuous} {Finite}
  {Element} {Method} for the {Numerical} {Solution} of the {Compressible}
  {Navier}–{Stokes} {Equations}},
\newblock \bibinfo{journal}{Journal of Computational Physics}
  \bibinfo{volume}{131} (\bibinfo{year}{1997}) \bibinfo{pages}{267--279}.
  \DOIprefix\doi{10.1006/jcph.1996.5572}.
\bibitem[{Hindenlang(2014)}]{hindenlang_mesh_2014}
\bibinfo{author}{F.~Hindenlang}, \bibinfo{title}{Mesh curving techniques for
  high order parallel simulations on unstructured meshes}, \bibinfo{type}{{PhD}
  thesis}, University of Stuttgart, \bibinfo{year}{2014}.
\bibitem[{Kennedy et~al.(2000)Kennedy, Carpenter, and
  Lewis}]{kennedy_low-storage_2000}
\bibinfo{author}{C.~A. Kennedy}, \bibinfo{author}{M.~H. Carpenter},
  \bibinfo{author}{R.~Lewis},
\newblock \bibinfo{title}{Low-storage, explicit {Runge}-{Kutta} schemes for the
  compressible {Navier}-{Stokes} equations},
\newblock \bibinfo{journal}{Applied Numerical Mathematics} \bibinfo{volume}{35}
  (\bibinfo{year}{2000}) \bibinfo{pages}{177--219}.
  \DOIprefix\doi{10.1016/S0168-9274(99)00141-5}.
\bibitem[{Kennedy and Carpenter(2003)}]{kennedy_additive_2003}
\bibinfo{author}{C.~A. Kennedy}, \bibinfo{author}{M.~H. Carpenter},
\newblock \bibinfo{title}{Additive {Runge}–{Kutta} schemes for
  convection–diffusion–reaction equations},
\newblock \bibinfo{journal}{Applied Numerical Mathematics} \bibinfo{volume}{44}
  (\bibinfo{year}{2003}) \bibinfo{pages}{139 -- 181}.
  \DOIprefix\doi{10.1016/S0168-9274(02)00138-1}.
\bibitem[{Vangelatos and Munz(2018)}]{vangelatos_study_2018}
\bibinfo{author}{S.~Vangelatos}, \bibinfo{author}{C.-D. Munz},
\newblock \bibinfo{title}{A {Study} on the {Performance} of {Implicit} {Time}
  {Integration} for the {Navier}-{Stokes} {Equations} using the {Discontinuous}
  {Galerkin} {Spectral} {Element} {Method}},
\newblock in: \bibinfo{booktitle}{Tenth {International} {Conference} on
  {Computational} {Fluid} {Dynamics}}, \bibinfo{publisher}{ICCFD 10},
  \bibinfo{address}{Barcelona}, \bibinfo{year}{2018}.
\bibitem[{Gassner(2009)}]{gassner_discontinuous_2009}
\bibinfo{author}{G.~Gassner}, \bibinfo{title}{Discontinuous {Galerkin}
  {Methods} for the {Unsteady} {Compressible} {Navier}-{Stokes} {Equations}},
  \bibinfo{type}{{PhD} {Thesis}}, University of Stuttgart,
  \bibinfo{address}{Stuttgart}, \bibinfo{year}{2009}.
\bibitem[{Blazek(2005)}]{blazek_computational_2005}
\bibinfo{author}{J.~Blazek}, \bibinfo{title}{Computational {Fluid} {Dynamics}:
  {Principles} and {Applications}}, \bibinfo{edition}{2nd} ed.,
  \bibinfo{publisher}{Elsevier Science}, \bibinfo{year}{2005}.
\bibitem[{Persson and Peraire(2006)}]{persson_sub-cell_2006}
\bibinfo{author}{P.-O. Persson}, \bibinfo{author}{J.~Peraire},
\newblock \bibinfo{title}{Sub-{Cell} {Shock} {Capturing} for {Discontinuous}
  {Galerkin} {Methods}},
\newblock in: \bibinfo{booktitle}{Proceedings of the 44th {AIAA} {Aerospace}
  {Sciences} {Meeting} and {Exhibit}}, \bibinfo{publisher}{American Institute
  of Aeronautics and Astronautics}, \bibinfo{address}{Reno},
  \bibinfo{year}{2006}.

\end{thebibliography}
